\documentclass[a4paper,13pt]{article}

\usepackage{epsfig}
\usepackage{color}
\usepackage{epsfig}
\usepackage[margin=2.75cm]{geometry}
\usepackage{epstopdf}
\usepackage{amsmath}\usepackage{amssymb}
\usepackage{mathrsfs}
\usepackage{graphicx}

\usepackage[colorlinks = true,
linkcolor = blue,
urlcolor  = blue,
citecolor = red,
anchorcolor = blue]{hyperref}

\begin{document}
\def \a'{\alpha'}
\baselineskip 0.60 cm
\begin{flushright}
\ \today
\end{flushright}

\begin{center}{\large
{\bf Scale-Invariant Two Component Dark Matter}} {\vskip 0.5 cm} {\bf ${\rm Seyed~ Yaser~ Ayazi}$$^1$ and ${\rm Ahmad~ Mohamadnejad}$$^{2}$}{\vskip 0.5 cm
}
{\small $^1$$Physics~Department$,~ $Semnan~University$,~$P.O.~Box.~35131-19111,~Semnan$,~$Iran$
$^2$$Young~Researchers~and~Elite~ Club, Islamshahr~ Branch, Islamic~ Azad~ University,$ $~ Islamshahr~ 3314767653, Iran$}

\end{center}

\begin{abstract}
We study a scale invariant extension of the standard model which can explain simultaneously dark matter and the hierarchy problem. In our set-up, we introduce  a scalar and a spinor as two-component dark matter in addition to scalon field as a mediator. Interesting point about our model is that due to scale invariant conditions, compared to other two-component dark matter models, it has lower independent parameters. Possible astrophysical and laboratory signatures of two-component dark matter candidate are explored and it is shown that the most contribution of observed relic density of dark matter can be determined by spinor dark matter.  Detectability of these dark matter particles is studied and the direct and invisible Higgs decay experiments are used to rule out part of the  parameter space of the model. In addition, the dark matter self-interactions are considered and shown that their contribution saturate this constraint in the resonant regions.
\end{abstract}

\section{Introduction} \label{sec1}
The standard Model (SM) has been established by the discovery of the Higgs boson  and it can explain almost all of experimental results obtained until now. However there are a number of unanswered issues, either theoretical or experimental such as hierarchy problem, active neutrino masses, dark matter (DM) relic abundance, baryon asymmetry of the Universe, inflation in the early Universe, dark energy, and so on. 

The existence of DM is inferred  through crucial
evidence such as galactic rotation curves, gravitational lensing,
observations of merging galaxies, the cosmic microwave background (CMB) measurements, the large scale structure of the universe and
the collision of the bullet clusters.   As it is mentioned,  there are still lack of experimental or observational evidences to precisely distinguish the correct particle physics model for DM physics. 

To explain these issues  a number of SM extensions such as supersymmetric standard model, technicolor and extra dimensions theories have been proposed. Despite the broad searches on the beyond SM at LHC, null results for beyond SM theories \cite{Susy result} shows that we have enough motivation to think about the alternative theories. 

In almost all extended models, there are some additional particles, which usually have heavier masses compared to the electroweak (EW) scale.
It is famous that the hierarchy problem arises from the fact that the negative Higgs mass term
in Lagrangian of SM causes a quadratical divergent term proportional to the energy scale cut-off $\Lambda^2$ after including the quantum corrections. 
As an idea avoiding the hierarchy problem, classically scale invariant extensions provides an attractive framework \cite{Gildener:1976ih}-\cite{Coleman:1973jx}. In this picture, it is supposed that
the tree-level Higgs mass is zero and in the quantum level the Higgs scalar gains a small mass from the radiative corrections. In fact, the Higgs mass term is the only term that breaks the classical scale invariance in the SM. Therefore by omitting the Higgs mass term from the SM potential, one can practically remove the hierarchy problem. In recent years, a lot of classically scale invariant models have been studied for the solution of the hierarchy problem  and DM problem \cite{scale invariant}. The possibility of other two-component models without concerning scale invariance have been extensively considered in literature \cite{two-component}. Also the two-component DM has been studied  in the context of scalar WIMP-like candidates \cite{ghorbani}.  Our goal in this paper is to address DM relic density and hierarchy problem by an extension of the scale invariant standard model (SISM) which contains a scalar and a spinor DM candidates. 

The structure of this paper is as follows: in section 2, we introduce the
scale invariant SM with two-component scalar and fermionic DM
scenarios. In section 3, we study perturbativity constraints on
two component scale-invariant DM. In section 4, we study
freeze-out solutions to the relic density constraint. In section
5, we will study phenomenological aspects such as direct
detection, indirect detection, self-interaction and invisible Higgs decay searches on
parameters space of our model. The results are summarized in
section 6. The decay rate and cross section formula for self-interaction of two component of DM are summarized in the appendix.

\section{The model}
In the SISM,  before electroweak symmetry breaking all fields in the scale invariant sector of potential are massless. In the quantum level these fields gain mass from radiative Coleman-Weinberg symmetry breaking \cite{Coleman:1973jx}. 

In this paper, we consider a scale-invariant extension of SM where Higgs mass term is absent, and the only term remaining in the Higgs potential will be $\lambda_{H} (H^{\dagger}H)^{2}$. In order to have a scale invariant version of the SM possessing a Higgs doublet and other SM
particles with their physical masses, at least two more
scalars (singlet) must be added to the theory. The reason arises from this fact that in the absent of scalar DM, the square scalon mass was completely fixed  and would be negative \cite{Gildener:1976ih}. In order to satisfy this condition, we add three new fields, two scalars and one spinor in our model. All fields are  singlets  under SM gauge transformation
and they are massless before spontaneous symmetry breaking. Two of these new fields, the scalar $ S $ and the spinor $ \chi $,
are assumed to be odd under a $ Z_{2} $ symmetry. These discrete symmetry
guarantees the stability of the lightest odd particles. The other scalar field, $ \phi $, and all SM particles are even under the $ Z_{2} $. Therefore under $ Z_{2} $ symmetry new fields transform as below:
\begin{equation} \label{1}
\phi \rightarrow \phi, \, \, S \rightarrow -S , \, \, \chi \rightarrow - \chi.
\end{equation}
The scalar part of the Lagrangian including the new fields is
\begin{equation} \label{2}
{\cal L}_{scalar} = \frac{1}{2} \partial_{\mu} \phi \,  \partial^{\mu} \phi
+ \frac{1}{2} \partial_{\mu} S \,  \partial^{\mu} S
+  D_{\mu} H^{\dagger}  \,  D^{\mu} H - V(H,\phi,S) ,
\end{equation}
where the most general scale-invariant potential $ V(H,\phi,S) $ which is renormalizable and invariant under gauge and $ Z_{2} $-symmetry is
\begin{align} \label{3}
V(H,\phi,S) & =  \frac{1}{6} \lambda_{H} (H^{\dagger}H)^{2}
+ \frac{1}{4!} \lambda_{\phi} \phi^{4} + \frac{1}{4!} \lambda_{s} S^{4} \nonumber \\
& + \lambda_{\phi H} \phi^{2}  H^{\dagger}H +
+ \lambda_{s H} S^{2} H^{\dagger}H + \lambda_{\phi s} \phi^{2}  S^{2}
\end{align}
where $H$, $\phi$ and $S$ are  the doublet Higgs, the scalon and DM scalars, respectively.

The scale-invariant terms including new spinor field and its allowed interaction are given by
\begin{equation} \label{4}
{\cal L}_{spinor} = \overline{\chi} ( i \gamma^{\mu} \partial_{\mu} - g \, \phi ) \chi .
\end{equation}

Since there are no allowed interaction terms in the Lagrangian
including both odd fields, the heavier odd particle also turns out to be stable.
Therefore, the model has an accidental symmetry that stabilizes the heavier odd particles
and it contains two DM candidates.

In unitary gauge, $ H = \frac{1}{\sqrt{2}} ( \begin{smallmatrix}
0 \\ h
\end{smallmatrix} ) $, potential (\ref{3}) becomes:
\begin{align} \label{5a}
V(h,\phi,S) & =  \frac{1}{4!} \lambda_{H} h^{4}
+ \frac{1}{4!} \lambda_{\phi} \phi^{4} + \frac{1}{4!} \lambda_{s} S^{4} \nonumber \\
& + \frac{1}{2} \lambda_{\phi H} \phi^{2}  h^{2} +
+ \frac{1}{2} \lambda_{s H} S^{2} h^{2} + \lambda_{\phi s} \phi^{2}  S^{2}
\end{align}
Minimum of potential (\ref{5a}) corresponds to fields vacuum expectation values. Necessary conditions for local minimum of $ V(h,\phi,S) $ are:
\begin{align} \label{5b}
\frac{\partial V}{\partial h} & = 0 \quad \Rightarrow \quad \dfrac{1}{3!} \lambda_{H} h^{3} + \lambda_{\phi H} \phi^{2}  h + \lambda_{s H} S^{2} h = 0 \nonumber \\
\frac{\partial V}{\partial \phi} & = 0 \quad \Rightarrow \quad \dfrac{1}{3!} \lambda_{\phi} \phi^{3} + \lambda_{\phi H} \phi h^{2} + 2 \lambda_{\phi s} S^{2} \phi = 0 \nonumber \\
\frac{\partial V}{\partial S} & = 0 \quad \Rightarrow \quad \dfrac{1}{3!} \lambda_{s} S^{3} + \lambda_{s H} S h^{2} + 2 \lambda_{\phi s} \phi^{2} S = 0
\end{align}
Eqs. (\ref{5b}) should hold for the fields vacuum expectation values. Note that we require the non-vanishing vacuum expectation values for the fields $ h $ and $ \phi $ so the scalar field $S$  remains stable because of the $Z_2$ symmetry and thereby it can play the role of the DM. Therefore, we put $ S = 0 $ in Eqs. (\ref{5b}):
\begin{align} \label{5c}
& \dfrac{1}{3!} \lambda_{H} h^{3} + \lambda_{\phi H} \phi^{2}  h  = 0 \nonumber \\
& \dfrac{1}{3!} \lambda_{\phi} \phi^{3} + \lambda_{\phi H} \phi h^{2} = 0
\end{align}
We are looking for non trivial solution of (\ref{5c}) corresponding to non-vanishing vacuum expectation values for $ h $ and $ \phi $. For non-zero $ h $ and $ \phi $ , Eq. (\ref{5c}) leads to 
\begin{equation} \label{5d}
\begin{pmatrix} \dfrac{1}{3!} \lambda_{H}~~~  \lambda_{\phi H} \\\lambda_{\phi H}  ~~~~~\dfrac{1}{3!} \lambda_{\phi}
 \end{pmatrix}\begin{pmatrix}
h^{2} \\  \phi^{2}
\end{pmatrix} = 0 \quad \Rightarrow \quad 
\begin{vmatrix} \dfrac{1}{3!} \lambda_{H}~~~  \lambda_{\phi H} \\\lambda_{\phi H}  ~~~~~\dfrac{1}{3!} \lambda_{\phi}
 \end{vmatrix} = 0,
\end{equation}
or simply:
\begin{equation} \label{5}
\lambda_{H} \, \lambda_{\phi} = (3! \, \lambda_{\phi H} )^{2} .
\end{equation}
Note that according to condition (\ref{5d}), the minimum of the potential term $ V(h,\phi,S) $ corresponding to vacuum expectation values of the fields is zero.

The filed $ H $ breaks the electroweak symmetry with vacuum expectation value, $ \langle H \rangle = \frac{1}{\sqrt{2}} ( \begin{smallmatrix}
0 \\ \nu_{1}
\end{smallmatrix} ) $, where $ \nu_1 = 246 \, GeV $.
Thus the Higgs field after spontaneous symmetry breaking is given by:
\begin{equation} \label{6}
H = \frac{1}{\sqrt{2}} \begin{pmatrix}
0 \\ \nu_{1} + h_{1}
\end{pmatrix}.
\end{equation}
As it was mentioned, the field $\phi$ also acquire a vacuum expectation value,
\begin{equation} \label{7}
\phi = \nu_{2} + h_{2} .
\end{equation}

Notice that $h_{1}$ and
$ h_{2} $ mix with each other and can be rewritten by the
mass eigenstates $ H_{1} $ and $ H_{2} $ as
\begin{equation} \label{8}
\begin{pmatrix}
H_{1}\\H_{2}\end{pmatrix}
 =\begin{pmatrix} cos \alpha~~~  -sin \alpha \\sin \alpha  ~~~~~cos \alpha
 \end{pmatrix}\begin{pmatrix}
h_1 \\  h_{2}
\end{pmatrix},
\end{equation}
where $ \alpha $ is the mixing angle.
We identify $ H_{1} $ with the SM-like Higgs observed at the LHC with a mass of about $125 \,~\rm GeV$. 

After the symmetry breaking, we have the following constraints:
\begin{align} \label{9}
& \nu_{2} =  \frac{M_{\chi}}{g} , \nonumber \\
& sin \alpha =  \frac{\frac{\nu_{1}}{\nu_{2}}}{\sqrt{1+(\frac{\nu_{1}}{\nu_{2}})^{2}}} \nonumber \\
& M_{H_{2}} =  0 \nonumber \\
& \lambda_{H} =  \frac{3 M_{H_{1}}^{2}}{ \nu_{1}^{2}} cos^{2} \alpha  \nonumber  \\
& \lambda_{\phi} =  \frac{3 M_{H_{1}}^{2}}{ \nu_{2}^{2}} sin^{2} \alpha  \nonumber  \\
& \lambda_{\phi H} =  - \frac{ M_{H_{1}}^{2}}{2 \nu_{1} \nu_{2} } sin \alpha \, cos \alpha \nonumber  \\
&\lambda_{s H} =  \frac{M_{s}^{2} - 2 \lambda_{\phi s} \nu_{2}^{2}}{\nu_{1}^{2}}, 
\end{align}
where $ M_{s} $ and $ M_{\chi} $ are the masses of scalar and spinor DM after symmetry breaking, respectively. The $ H_{2} $ field (scalon) is massless, and it can be shown that the elastic scattering cross section of DM off nuclei becomes drastically large and the model is immediately excluded by the direct detection experiments. The one-loop correction gives a
mass to the  massless eigenstate $H_{2}$\cite{Gildener:1976ih},\cite{ghorbani}:
\begin{equation} \label{10}
M_{H_{2}}^2 =  - \frac{\lambda_{\phi H}}{16 \pi^{2} M_{H_1}^2 } (M_{H_1}^4 + M_{s}^4 + 6M_{W}^4 + 3M_{Z}^4 - 4M_{\chi}^4 - 12M_{t}^4) .
\end{equation}
Notice that in the absence of scalar and fermionic DM, scalon mass was completely fixed by Higgs particle,  the $Z$ gauge boson and the top quark masses. For this reason, adding scalar field is inevitable. Moreover in the absence of additional scalar DM, the square scalon mass could be negative. Since $ M_{H_{2}}^{2} > 0 $ and $ \lambda_{\phi H} < 0 $ , Equation (\ref{10}) leads to the following constraint on $ M_{s} $ :
\begin{align}
M_{s} > f(M_{\chi}) \label{11}
\end{align}
where 
\begin{align}
f(M_{\chi}) = \sqrt[4]{4M_{\chi}^4-(M_{H_{1}}^{4}+6M_{W}^{4}+3M_{Z}^{4}-12M_{t}^{4})}, \label{12}
\end{align}
and $ f(0) = 310.7 ~\rm  GeV $ which is the minimum of $M_{s}$.
Throughout this paper, we satisfy this condition.

According to (\ref{9}), the model introduces only 5 free parameters including
$ \lambda_{s} \, , \, \lambda_{\phi s} \, , \, M_{s} \, , \, M_{\chi} \, , \,  g  $.
In addition, the quartic coupling $ \lambda_{s} $ is
irrelevant to the DM relic density.
Therefore, the remaining free parameters are
\begin{equation} \label{13}
\lambda_{\phi s} \, , \, M_{s} \, , \, M_{\chi} \, , \,  g .
\end{equation}

It is remarkable that our model in comparison with other two-component DM models, has a much lower number of independent parameters and behaves like a single-component model. For this reason, it would be difficult to satisfy all theoretical and phenomenological constraints, simultaneously. This is the point that we encounter in the next sections. In the following, we examine perturbativity constraints on these four parameters.

\section{Theoretical constraints} \label{sec3}
In this section, we discuss various constraints on the parameters of our model from  theoretical considerations. These
are furnished in the following. Perturbativity constraints on the parameters of the Lagrangian are
\begin{align}
& - 4 \pi < \lambda_{H} \, , \, \lambda_{\phi} \, , \, \lambda_{s}
\, , \, \lambda_{\phi H} \, , \, \lambda_{s H} \, , \, g  < 4 \pi \label{31} \\
& - 8 \pi < \lambda_{\phi s} < 8 \pi  \label{32}
\end{align}
Considering constraints (\ref{9}) we have
\begin{align}
0 & < \frac{3 M_{H_{1}}^{2}}{ \nu_{1}^{2}} cos^{2} \alpha < 4 \pi \label{33} \\
0 & < \frac{3 M_{H_{1}}^{2}}{ \nu_{2}^{2}} sin^{2} \alpha  < 4 \pi \label{34} \\
0 & < \frac{ M_{H_{1}}^{2}}{2 \nu_{1} \nu_{2} } sin \alpha \, cos \alpha <  4 \pi \label{35} \\
- 4 \pi & < \frac{M_{s}^{2} - 2 \lambda_{\phi s} \nu_{2}^{2}}{\nu_{1}^{2}} < 4 \pi \label{36} \\
0 & < g <  4 \pi \label{37}
\end{align}
One can easily show that Eq.~(\ref{33}) and Eq.~(\ref{35}) are established automatically.
Constraint (\ref{34}) leads to
\begin{equation}
0 < sin \, \alpha < max(sin \alpha) \label{38}
\end{equation}
where $ max(sin \alpha) = \sqrt{\sqrt{A^{2}+2A}-A} $ with $ A = \frac{2 \pi \nu_{1}^{2}}{3 M_{H_{1}}^{2}} $ ($ max(sin \alpha) = 0.972 $).
The above equation, $ 0 < sin \, \alpha < 0.972 $, is not a strong constraint on $ sin \, \alpha $. However, it leads to a constraint on $ M_{\chi} $:
\begin{equation}
M_{\chi} > \frac{\sqrt{1-[max(sin \alpha)]^{2}}}{max(sin \alpha)} g \nu_{1}
= (59.38 \, GeV) \, g  \label{381}
\end{equation}

Regarding to Eq.~(\ref{36})
\begin{equation}
2 \lambda_{\phi s} \nu_{2}^{2} - 4 \pi \nu_{1}^{2}  < M_{s}^{2} <
2 \lambda_{\phi s} \nu_{2}^{2} + 4 \pi \nu_{1}^{2} , \label{39}
\end{equation}
and according to (\ref{11}) we have,
\begin{equation}
f^{2}(M_{\chi}) < M_{s}^{2} < 2 \lambda_{\phi s} \nu_{2}^{2} + 4 \pi \nu_{1}^{2} \, \Rightarrow \, (\frac{f^{2}(M_{\chi})}{2})^{2} < (\lambda_{\phi s} \nu_{2}^{2} + 2 \pi \nu_{1}^{2})^{2}.
 \label{310}
\end{equation}
Considering $ f^{4}(M_{\chi}) = 4 M_{\chi}^{4} + f^{4}(0) $ and $ \nu_{2} = \frac{M_{\chi}}{g} $, Eq.~(\ref{310}) leads to
\begin{equation}
a M_{\chi}^{4} - b M_{\chi}^{2} - c < 0,  \label{3102}
\end{equation}
where $ a = 1-\frac{\lambda_{\phi s}^{2}}{g^{4}} $, $ b = \frac{4 \pi \lambda_{\phi s} \nu_{1}^{2} }{g^{2}} $, and $ c = 4 \pi^{2} \nu_{1}^{4} 
- \frac{f^{4}(0)}{4} > 0 $. For $ \lambda_{\phi s} > 0 $ ($ b > 0 $), there are two possibilities: first  $ a < 0 $, so Eq.~(\ref{3102}) is trivial and second $ a > 0 $, provides a constraint on $ M_{\chi} $:
\begin{equation}
M_{\chi} < \sqrt{\frac{b+\sqrt{b^{2}+4ac}}{2a}},  \label{3103}
\end{equation}

Finally, we choose the following domains for the parameters space: (\ref{13})
\begin{align}
& 0 < g < 4 \pi  \label{311} \\
& 0 < \lambda_{\phi s}  < 8 \pi  \label{312} \\
& (59.38 \,~\rm {GeV}) \, g  < M_{\chi} < \sqrt{\frac{b+\sqrt{b^{2}+4|a|c}}{2|a|}} \label{313} \\
& \sqrt{max(f^{2}(M_{\chi}) \, , \,2 \lambda_{\phi s} \nu_{2}^{2} - 4 \pi \nu_{1}^{2})} < M_{s} < \sqrt{ 2 \lambda_{\phi s} \nu_{2}^{2} + 4 \pi \nu_{1}^{2} } \label{314}
\end{align}

\section{Relic abundance} \label{sec4}
The evolution of the number density of DM particles with time are governed by the Boltzmann equation. 
In this section, we compute the relic density for both 
DM candidates scalar and fermion in our model, at the present epoch. In general, the coupled Boltzmann equations for two-component DM $S$ and $\chi$
should be solved in order to compute the number density. The coupled Boltzmann equations for scalar $S$ and fermion $\chi$ are given by:
\begin{eqnarray}
\frac{d n_{\chi}}{dt}+3Hn_{\chi}&=&-\sum_{j=p,H_1,H_2} \langle\sigma_{\chi\chi\rightarrow jj}\upsilon\rangle (n^2_{\chi}-n^2_{\chi,eq})\nonumber\\&&-\langle\sigma_{\chi\chi\rightarrow SS}\upsilon\rangle (n^2_{\chi}-n^2_{\chi,eq}\frac{n^2_{S}}{n^2_{S,eq}})
\label{Boltzman-X},
\end{eqnarray}
\begin{eqnarray}
\frac{d n_{S}}{dt}+3Hn_{S}&=&-\sum_{j=p,H_1,H_2} \langle\sigma_{SS\rightarrow jj}\upsilon\rangle (n^2_{S}-n^2_{S,eq})\nonumber\\&&-\langle\sigma_{SS\rightarrow \chi\chi}\upsilon\rangle (n^2_{S}-n^2_{S,eq}\frac{n^2_{\chi}}{n^2_{\chi,eq}})
\label{Boltzman-S},
\end{eqnarray}
where $p$ denotes any SM particles. In $\langle\sigma_{ab\rightarrow cd}\upsilon\rangle$  all annihilations  are taken into account except $\langle\sigma_{S\chi\rightarrow S\chi}\upsilon\rangle$ which does not affect the number density. By using $x = m/T$, where $T$ is the photon temperature, as the independent variable instead of time and $\dot{T}=-HT$, one can rewrite the Boltzmann equations in terms of yield quantity, $Y=n/s$:
\begin{eqnarray}
\frac{dY_{\chi}}{dx}&=&-\sqrt{\frac{45}{\pi}} M_{pl} \, g_{*}^{1/2} \, \frac{m}{x^{2}}
[\sum_{j=p,H_1,H_2}\langle\sigma_{\chi \chi \rightarrow jj}v\rangle(Y_{\chi}^{2}-Y_{\chi,eq}^{2})\nonumber\\&&+
\langle\sigma_{\chi \chi \rightarrow SS}v\rangle(Y_{\chi}^{2}-Y_{\chi,eq}^{2} \frac{Y_{S}^{2}}{Y_{S,eq}^{2}})], \label{41} 
\end{eqnarray}
\begin{eqnarray}
\frac{dY_{S}}{dx} &=& - \sqrt{\frac{45}{\pi}} M_{pl} \, g_{*}^{1/2} \, \frac{m}{x^{2}}
[\sum_{j=p,H_1,H_2}\langle\sigma_{SS \rightarrow jj }v\rangle(Y_{S}^{2}-Y_{S,eq}^{2}) \nonumber\\&&+
\langle\sigma_{SS \rightarrow \chi \chi}v\rangle(Y_{S}^{2}-Y_{S,eq}^{2} \frac{Y_{\chi}^{2}}{Y_{\chi,eq}^{2}})], \label{42}
\end{eqnarray}
where $M_{pl}$ is the Planck mass and $g_{*}^{1/2}$ is the effective numbers parameter. As it is seen in above equations, there are new terms in Boltzmann equations which describe the conversion of two DM particles into each other, $ \langle SS \leftrightarrow \chi \chi \rangle$. 
These two cross sections are also described by the same matrix element. Therefore, we expect $ \langle\sigma_{\chi \chi \rightarrow SS}v \rangle$ and $\langle \sigma_{SS \rightarrow \chi \chi}v\rangle $ are not independent and their relation is:
\begin{equation} \label{43}
Y_{\chi,eq}^{2} \langle\sigma _{\chi \chi \rightarrow SS}v\rangle =
Y_{S,eq}^{2} \langle\sigma _{SS \rightarrow \chi \chi}v\rangle .
\end{equation}

The interactions between the two DM components take place by exchanging two scalar mass eigenstates $H_1$ and $H_2$ where the coupling of $\chi $ to $ H_{1} $ is suppressed by $ sin \, \alpha $. Therefore,
it usually is the $ H_{2} $-mediated diagram that gives the dominant contribution.
However, If one DM particle is heavier than the other one (\ref{12}), the conversion of the heavier particle into the lighter one is relevant, $ SS \rightarrow \chi \chi $. Thus, the contribution of $ \chi $ in the relic density is dominant and the only option for annihilation of $ \chi $ is via $ H_{1} $-mediated and $ H_{2} $-mediated diagrams into SM particles. 

To solve numerically the two coupled Boltzmann differential equation, We have implemented the model into micrOMEGAs \cite{Belanger:2014vza} (via LanHEP \cite{Semenov:2008jy}). Since we have two stable DM particle, the DM constraint in this model reads
\begin{equation} \label{44}
\Omega_{DM} h^{2} = \Omega_{S} h^{2} + \Omega_{\chi} h^{2} = 0.1199 \pm 0.0027
\end{equation}
according to the data by Planck collaboration \cite{Planck}.
Another related quantity is the fraction of the DM density that is due to $ S $ and $ \chi $
 denoted by $ \xi_{S} $ and $ \xi_{\chi} $,  respectively. So
\begin{equation} \label{45}
 \xi_{\chi} = \frac{\Omega_{\chi}}{\Omega_{DM}}, \quad
  \xi_{S} = \frac{\Omega_{S}}{\Omega_{DM}}, \quad \rm with \, \, \, \xi_{\chi}  + \xi_{S}  = 1.
\end{equation}

\begin{figure}[!htb]
\centerline{\hspace{0cm}\epsfig{figure=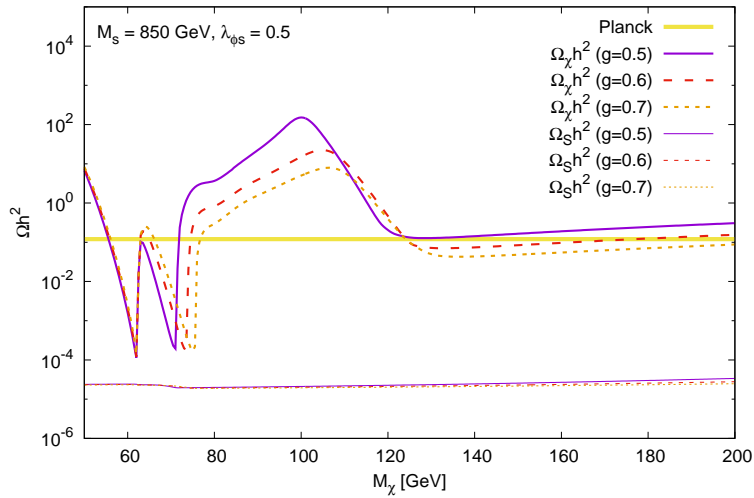,width=10cm}}
\caption{Relic density as a function of fermionic DM mass for differnet values of coupling $ g $.}\label{Relic Density1}
\end{figure}

\begin{figure}[!htb]
\centerline{\hspace{0cm}\epsfig{figure=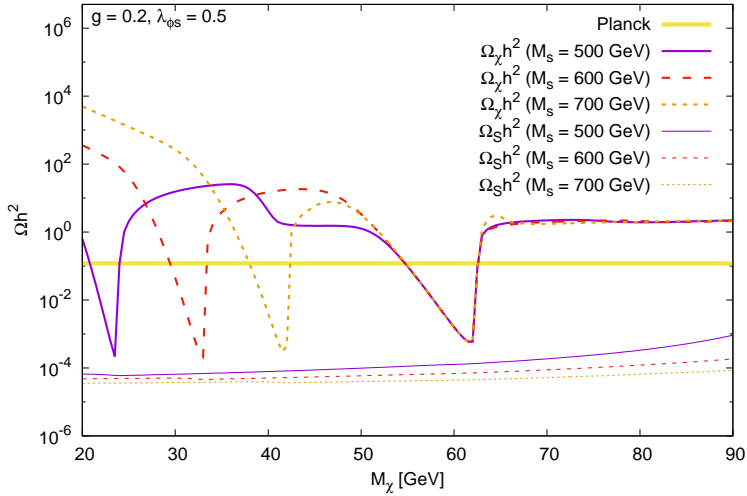,width=10cm}}
\caption{Relic density as a function of fermionic DM mass for differnet values of scalar mass $ M_{s} $.}\label{Relic Density2}
\end{figure}

\begin{figure}[!htb]
\centerline{\hspace{0cm}\epsfig{figure=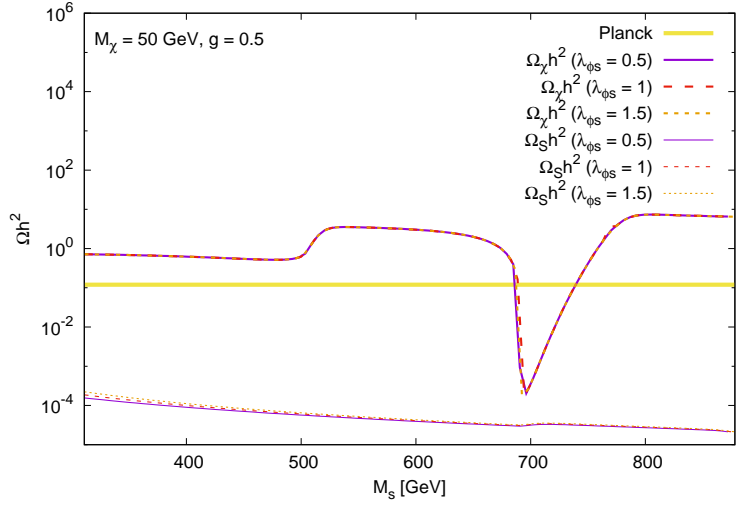,width=10cm}}
\caption{Relic density as a function of scalar DM mass for differnet values of coupling $ \lambda_{\phi s} $.}\label{Relic Density3}
\end{figure}

\begin{figure}[!htb]
\centerline{\hspace{0cm}\epsfig{figure=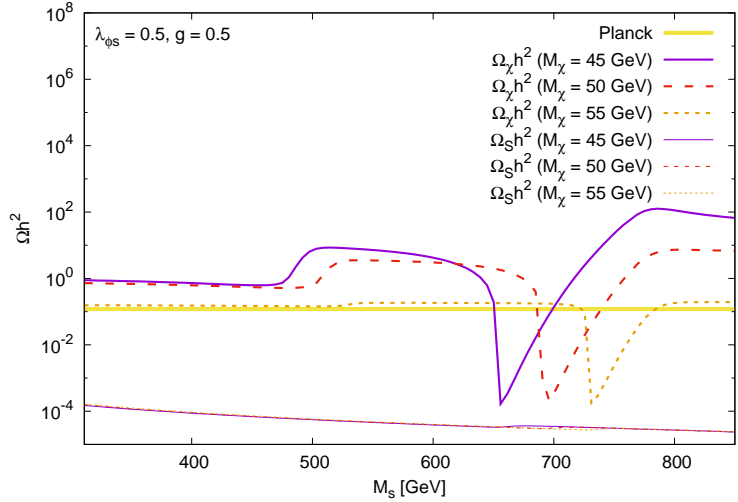,width=10cm}}
\caption{Relic density as a function of scalar DM mass for differnet values of fermionic mass $ M_{\chi} $.}\label{Relic Density4}
\end{figure}

Fig.~\ref{Relic Density1} to~\ref{Relic Density4} depict the relic density of fermionic and scalar DM as a function of the DM mass. According to these plots, the most contribution of DM relic density $ \Omega_{DM} $ comes from fermionic DM, i.e., $ \Omega_{\chi} $. Since, in our model scalar DM is always heavier than fermionic DM, in addition to annihilation to SM particles, it could also annihilate to fermionic DM particles. Therefore, its relic density is smaller than fermionic relic density.

Fig.~\ref{Relic Density1} shows both DM
relic densities as a function of $ M_{\chi} $ for different values of $ g $. For any given value
of $ g $ the fermionic relic density $ \Omega_{\chi} $ features a double reduction at the $ H_{1} $ and $ H_{2} $ resonances (respectively at $ M_{\chi} = \frac{M_{H_{1}}}{2} = 62.5 ~\rm GeV $, and $ M_{\chi} = \frac{M_{H_{2}}}{2} $). There is another reduction due to the opening of the $ \chi \chi \rightarrow H_{2} H_{2}  $ annihilation channel. Note that, according to Eq. (\ref{10}), $ M_{H_{2}} $ itself depends on $ g $, $ M_{s} $ and $ M_{\chi} $, so it is not an independent parameter. Therefore, in our relic density plots, it varies with $ g $ and DM masses.
In Fig.~\ref{Relic Density1}, scalar relic density $ \Omega_{S} $ does not vary dramatically with $ M_{\chi} $ or $ g $. Note that $ \lambda_{s H} $ is a determinative parameter in scalar DM annihilation to SM particles. On the other hand, annihilation of scalar DM to SM particles is more favorable than its annihilation to fermionic DM, because most SM particles are lighter than fermionic DM. Therefore, $ \Omega_{S} $ mostly depends on $ \lambda_{s H} $ rather than $ \lambda_{\phi s} $. 
According to Eqs. (\ref{9}) $ \lambda_{s H} $ is given by
\begin{equation} \label{neweq}
\lambda_{s H} =  \frac{M_{s}^{2}}{\nu_{1}^{2}} - \frac{2 \lambda_{\phi s} M_{\chi}^{2}}{g^{2} \nu_{1}^{2}},
\end{equation}
and for the given parameters in Fig.~\ref{Relic Density1}, $ \lambda_{s H} $ is mostly determined by the first term of Eq. (\ref{neweq}). Thus, it does not vary much with $ M_{\chi} $ or $ g $.

In Fig.~\ref{Relic Density2} DM relic densities are plotted versus $ M_{\chi} $ for different values of $ M_{s} $. Similarly, for the given values
of $ M_{s} $ the fermionic relic density again features a double reduction at the $ H_{2} $ and $ H_{1} $ resonances (respectively
at $ M_{\chi} = \frac{M_{H_{2}}}{2} $, and $ M_{\chi} = \frac{M_{H_{1}}}{2} = 62.5 \, GeV $). Obviously, in this plot $ M_{H_{2}} $ at the first resonance is lighter than $ M_{H_{1}} = 125 \, GeV $. For the scalar relic density, according to Eq. (\ref{neweq}), larger $ M_{s} $ leads to larger $ \lambda_{s H} $ and therefore DM-SM interaction gets stronger which leads to smaller scalar relic density. Furthermore, now for $ M_{s} = 500 ~\rm GeV $, the second term of Eq. (\ref{neweq}) can compete with the first term, and with growth of $ M_{\chi} $, $ \lambda_{s H} $ will decrease. Due to this reduction, scalar DM-SM interaction becomes weaker and therefore $ \Omega_{S} $ increase with $ M_{\chi} $. For larger $ M_{s} $ (for example $ M_{s} = 700 ~\rm GeV $ again the first term of Eq. (\ref{neweq}) dominates and $ \Omega_{S} $ increases less with $ M_{\chi} $.

Fig.~\ref{Relic Density3} and~\ref{Relic Density4} depict relic densities versus $ M_{s} $. In Fig.~\ref{Relic Density3}, for $ M_{\chi} = 50 ~\rm GeV $ there is a single reduction in fermionic relic density around $ M_{s} = 700 ~\rm GeV $. This reduction corresponds to $ M_{H_{2}} = 2 M_{\chi} = 100 ~\rm GeV $ which is a resonance case. According to Eq. (\ref{neweq}), $ \lambda_{s H} $ increases with $ M_{s} $ and scalar DM-SM interaction becomes stronger. Therefore, $ \Omega_{S} $ decrease with $ M_{s} $. In addition, for the given parameters, since the first term of Eq. (\ref{neweq}) dominates, $ \lambda_{s H} $ and therefore $ \Omega_{S} $ is nearly independent of $ \lambda_{\phi s} $. In this figure, only for small $ M_{s} $ a little dependency of $ \Omega_{S} $ to $ \lambda_{\phi s} $ can be realized.

Finally, in Fig.~\ref{Relic Density4} we display fermionic relic density as a function of $M_s$ for different values of $ M_{\chi} $. Therefore, we have different resonance cases corresponding to $ M_{H_{2}} = 2 M_{\chi} $ for each value of $ M_{\chi} $. For the given parameters, scalar relic density is not sensitive to different values of $ M_{\chi} $, because as it was mentioned before $ \Omega_{S} $ is mostly determined by $ \lambda_{s H} $ which again according to Eq. (\ref{neweq}), 
the second term can be neglected in comparison with the first term. Thus, for the given values of Fig.~\ref{Relic Density4}, only first term which is independent of $ M_{\chi} $ affects scalar DM relic density so that by growth of $ M_{s} $, $ \lambda_{s H} $ increases and consequently $ \Omega_{S} $ decreases.

In our model, total DM relic density does not depend on the $ \lambda_{\phi s} $. This parameter can only affect $ \Omega_{S} $ which has a small contribution in $ \Omega_{DM} = \Omega_{S} + \Omega_{\chi}$. Therefore, $ \Omega_{DM} $ only depends on $ g $, $ M_{s} $, and $ M_{\chi} $.

\section{Phenomenological aspects}
\subsection{Direct detection}

In this section, we investigate constraints on
parameters space of our model which are imposed by search for
scattering of DM-nuclei.  Since no such collision events have been observed yet by different DM direct detection experiments, these experiments provide an exclusion limit on DM-nucleon scattering cross-section. The strongest bounds on the DM-nucleon
cross section have been obtained by XENON100 \cite{XENON100} and LUX \cite{NewLUX} experiments.
\begin{eqnarray} \label{constraints}
\rm {XENON100}: \sigma_{SI}\leq 2\times10^{-45}~cm^2\nonumber\\ 
\rm{LUX} : \sigma_{SI}\leq 2.2\times10^{-46}~cm^2\nonumber 
\end{eqnarray}

\begin{figure}[!htb]
\begin{center}
\centerline{\hspace{0cm}\epsfig{figure=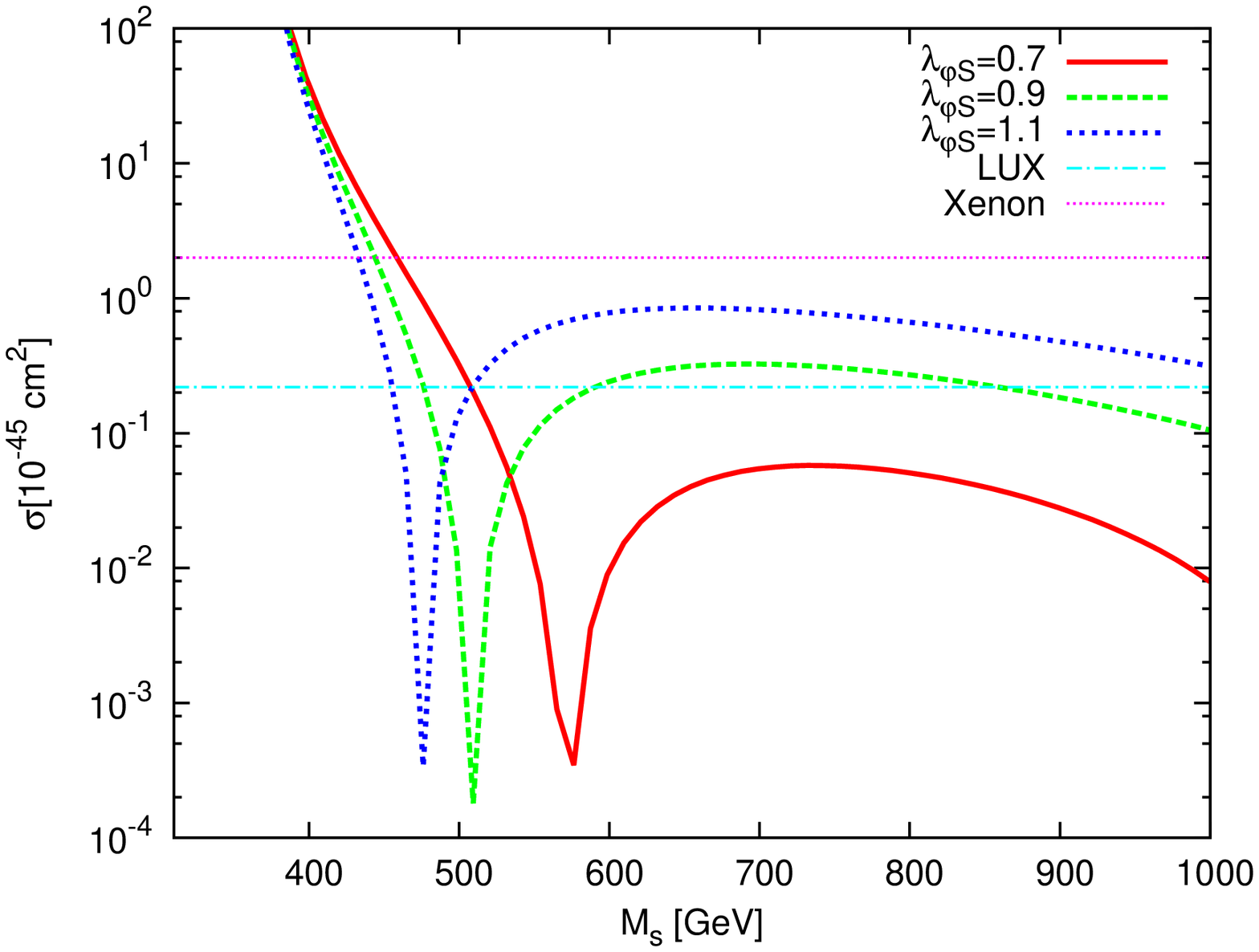,width=7cm}\hspace{0.3cm}\epsfig{figure=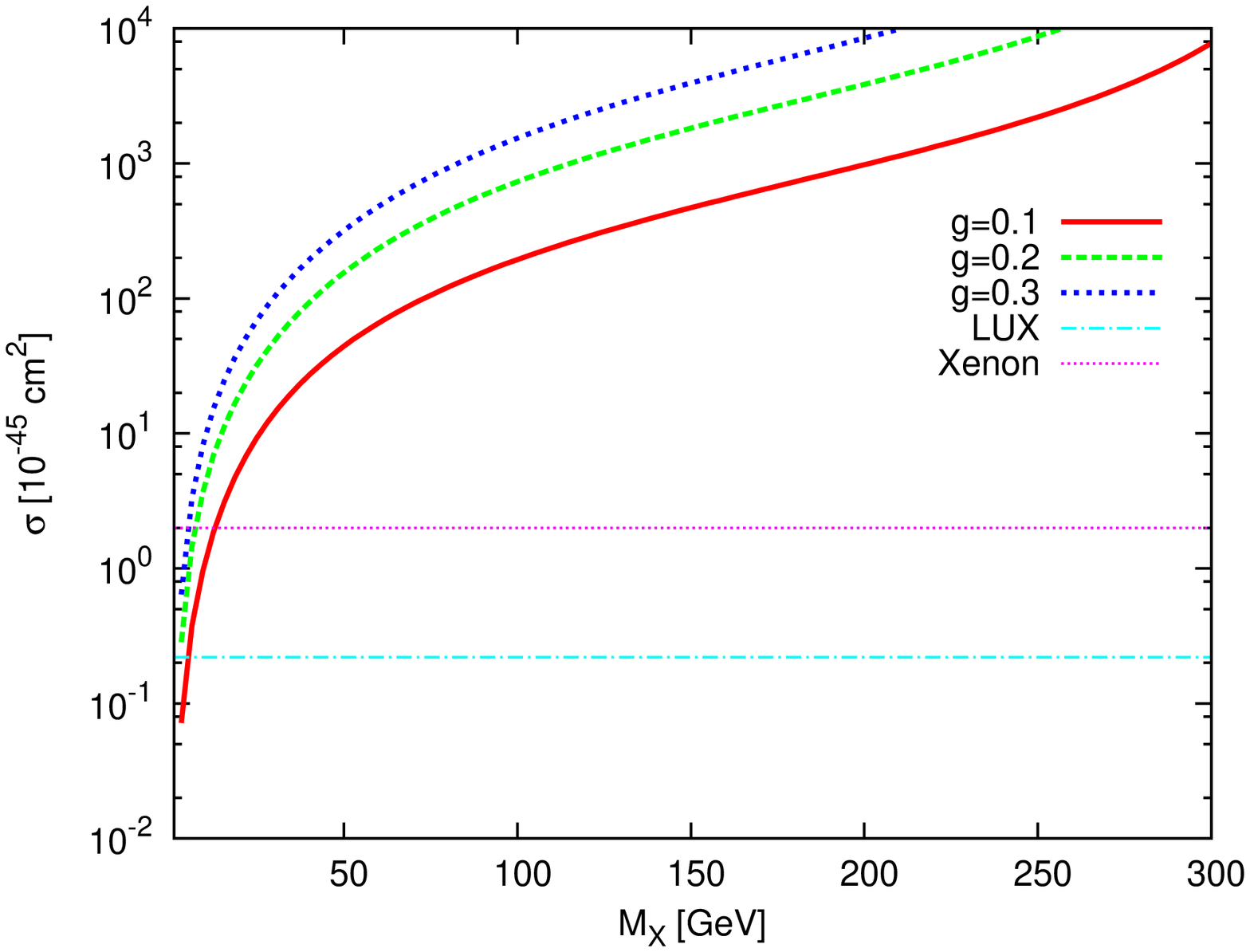,width=7cm}}
\centerline{\vspace{-1cm}\hspace{0.5cm}(a)\hspace{6cm}(b)}
\centerline{\vspace{-0.0cm}}
\end{center}
\caption{The direct detection cross section as a function of mass of (a) scalar DM.  We set  $M_{\chi}=200~\rm GeV$ and $g=0.2$.  (b) fermion DM. We set  $M_s=500~\rm GeV$.}\label{Direct}
\end{figure}

\begin{figure}[!htb]
\begin{center}
\centerline{\hspace{0cm}\epsfig{figure=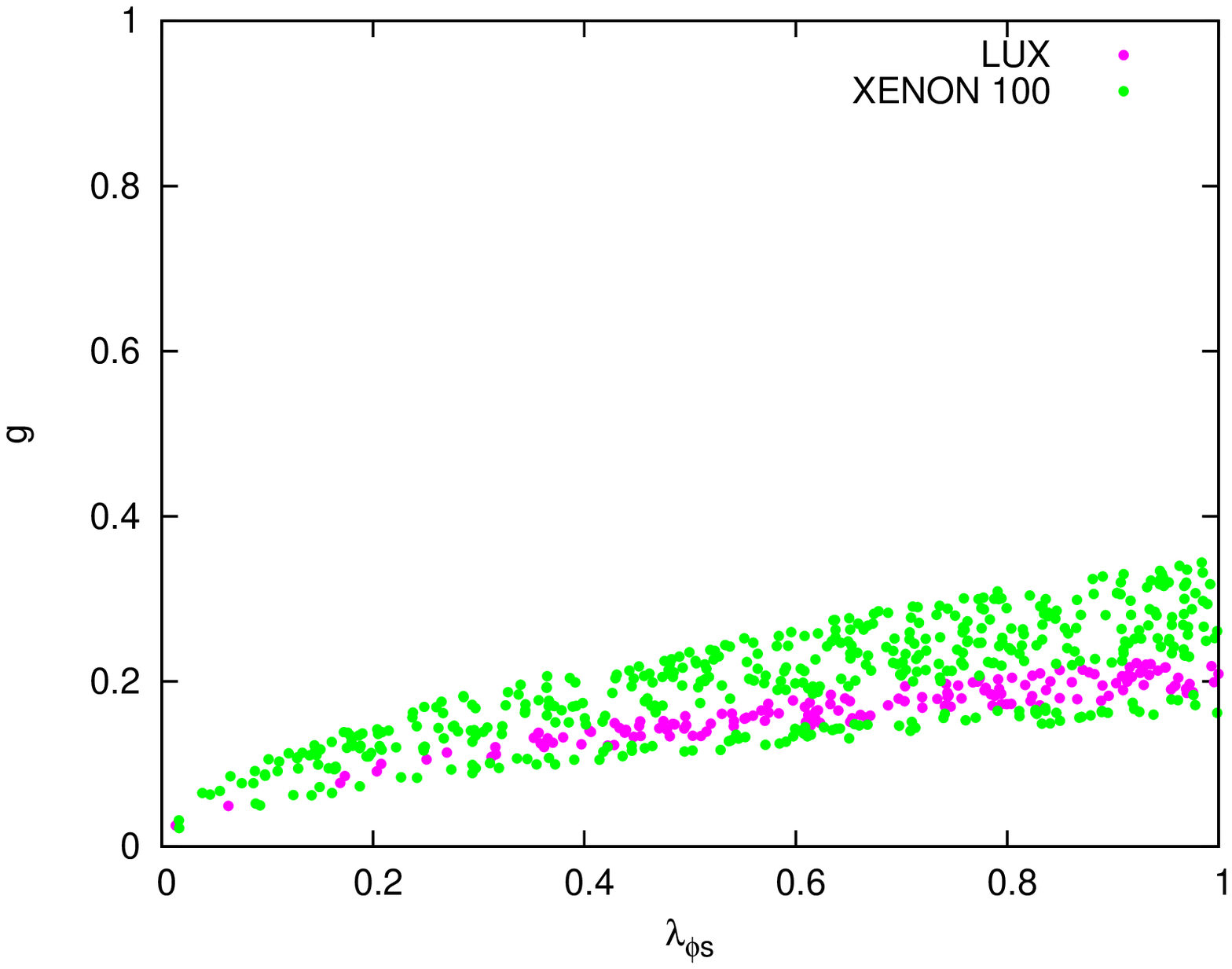,width=7.5cm}\hspace{-0.3cm}\epsfig{figure=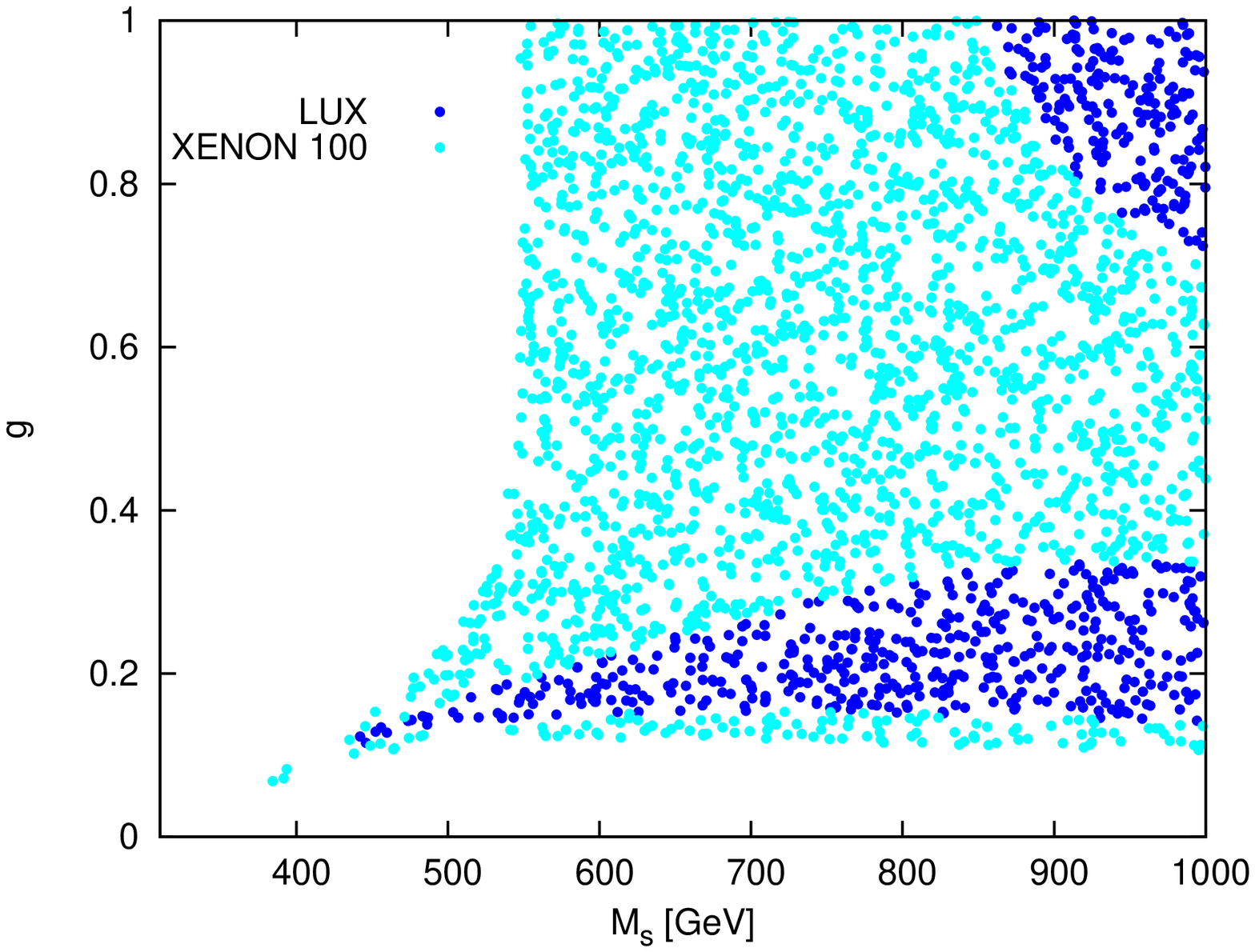,width=7.5cm}}
\centerline{\vspace{0.5cm}\hspace{0.5cm}(a)\hspace{6cm}(b)}
\centerline{\hspace{0cm}\epsfig{figure=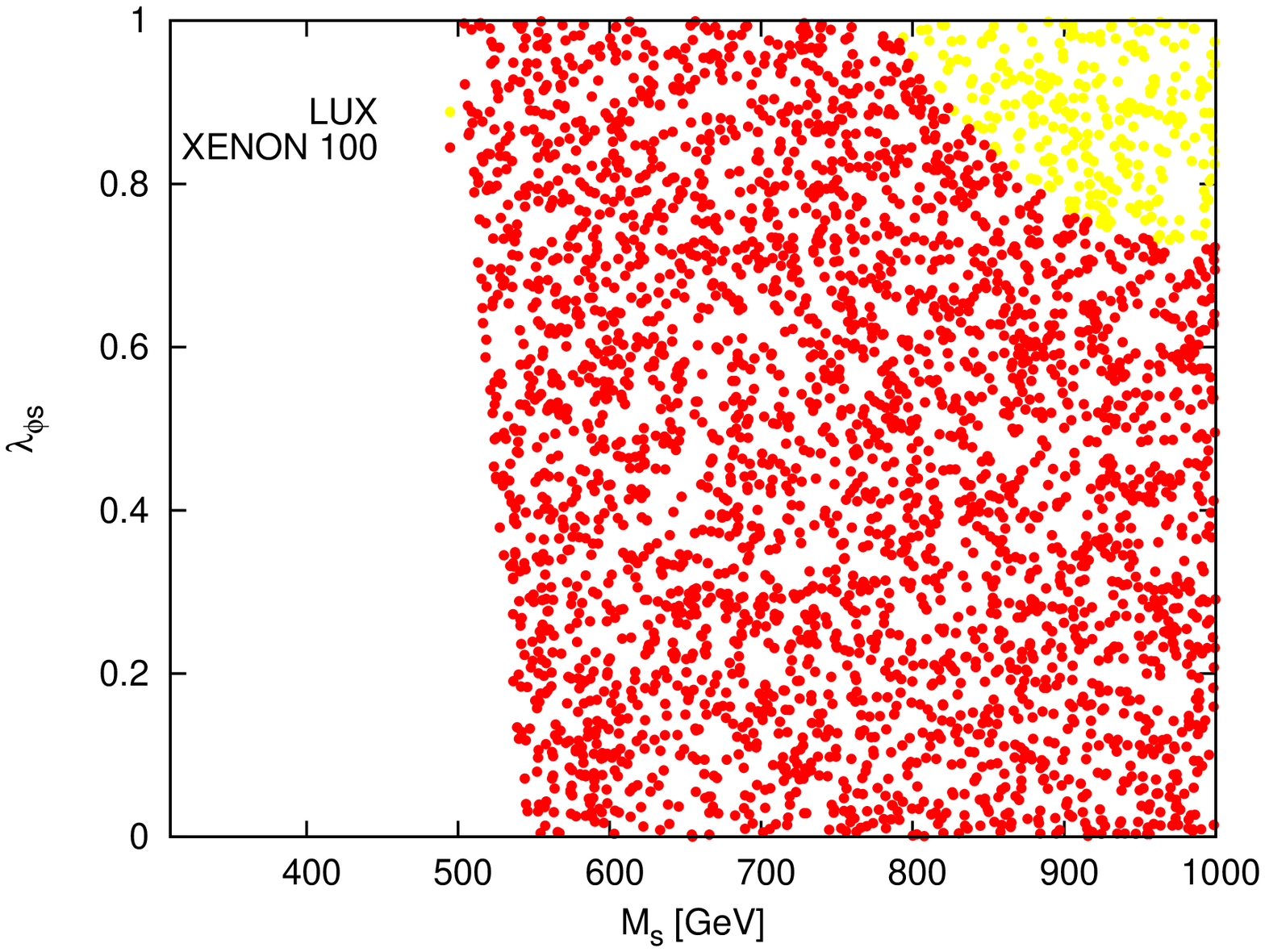,width=7.5cm}\hspace{-0.2cm}\epsfig{figure=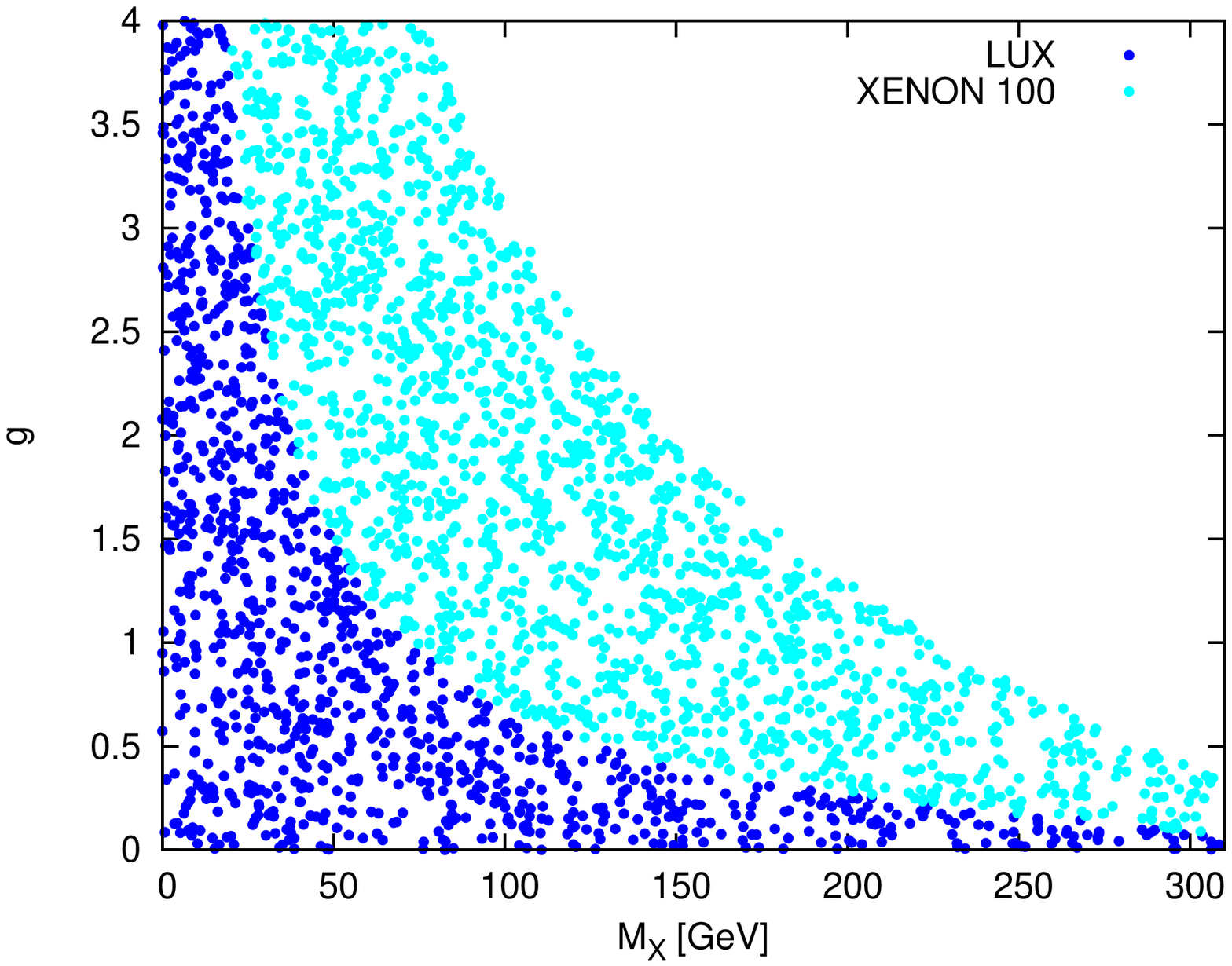,width=7.5cm}}
\centerline{\vspace{-1.2cm}\hspace{0.5cm}(c)\hspace{6cm}(d)}
\centerline{\vspace{-0.0cm}}
\end{center}
\caption{(a), (b) and (c) depict ranges of parameters space in  $g$, $\lambda_{\phi s}$ and $M_s$ planes for  scalar DM and (d) depicts allowed regions in $g$ and $M_{\chi}$  for fermionic DM  which are consistent with experimental measurements of $\sigma_{\rm Xenon100}$ and $\sigma_{\rm LUX}$. In (a), we set $M_s=500~\rm GeV$ and $M_{\chi}=200~\rm GeV$.  In (b), we set $M_{\chi}=200 ~\rm GeV$ and $g=0.5$. In (d), we set $M_s=500~\rm GeV$.}\label{scaterDirect}
\end{figure}

The spin-independent direct detection cross section of $\chi$ is determined by $ H_{1} $
and $ H_{2} $ exchanged diagrams:
 \begin{equation} \label{51}
 \sigma_{\chi} =  \xi_{\chi}\dfrac{g^3\nu_1}{\pi M_{\chi}(1+(\nu_1g/M_{\chi})^2)}\mu_{\chi}^2(\dfrac{1}{m^2_{H_1}}-\dfrac{1}{m^2_{H_2}})^2f_n^2
  \end{equation}
where $\xi_{\chi} = \frac{\Omega_{\chi}}{\Omega_{DM}}$ and $\mu_{\chi}$ is the reduced mass of nucleon and fermionic DM and
the coupling constant $f_n$ is given by nuclear matrix elements
and nucleon mass\cite{He:2008qm}. Similarly, for the scalar DM
candidate the effective spin independent direct detection
cross-section is given by:
\begin{align} \label{directScalar}
\sigma_{S}&= \xi_{S}\dfrac{\mu_S^2}{4\pi M^4_{H_1}M^4_{H_2}m^2_S}[\frac{M^2_s-2\lambda_{\phi s}M^2_{\chi}/g^2}{\nu_1}(\frac{M^2_{H_2}}{1+(\nu_1g/M_{\chi})^2}+\frac{M^2_{H_1}g^2\nu^2_1}{g^2\nu_1^2+M_{\chi}^2}) \nonumber \\
&+ \frac{2\nu_1\lambda_{\phi s}}{1+(\nu_1g/M_{\chi})^2}(M^2_{H_1}-M^2_{H_2})]^2f_n^2
\end{align}
where $\xi_{S} = \frac{\Omega_{S}}{\Omega_{DM}}$ and $\mu_S$  is the reduced mass of nucleon and scalar DM.
The parameters $\lambda_{\phi s}$ and $g$ are independent and have been defined in previous section. It is remarkable that the two terms  in Eq.~\ref{directScalar} may cancel against each other, giving a suppressed cross section. In Fig.~\ref{Direct}, we display the direct detection cross section as a function of mass of scalar and  fermion DM. As it is seen in Fig.~\ref{Direct}-a, $\sigma_S$ has a minimum in value of $ M_s$ which cancellation takes place. For scalar DM direct detection cross section depends to scalar DM mass, $\lambda_{\phi s}$, $g$ and $M_{\chi}$. While fermionic DM direct detection cross section does not depend to $\lambda_{\phi s}$. However as it is mentioned in previous section, $m_{H_2}$ is not an independent parameter and depends on three independent parameters of our model $M_s$, $g$ and $M_{\chi}$. Also $m_{H_2}$ may be very small and so the contribution of its propagator to the direct detection cross section can be very large. For this reason large portion of parameters space is excluded by this observable. In order to show allowed region in parameters space, we display  scatter points in Fig.~\ref{scaterDirect}. Figures.~\ref{scaterDirect}-a,b,c depict allowed regions in $g$, $\lambda_{\phi s}$ and $M_s$ for  scalar DM and Figure.~\ref{scaterDirect}-d depicts allowed regions in $g$ and $M_{\chi}$ for fermionic DM  which are consistent with
experimental measurements of $\sigma_{\rm Xenon100}$ and $\sigma_{\rm LUX}$. 

\begin{figure}[!htb]
\begin{center}
\centerline{\hspace{0cm}\epsfig{figure=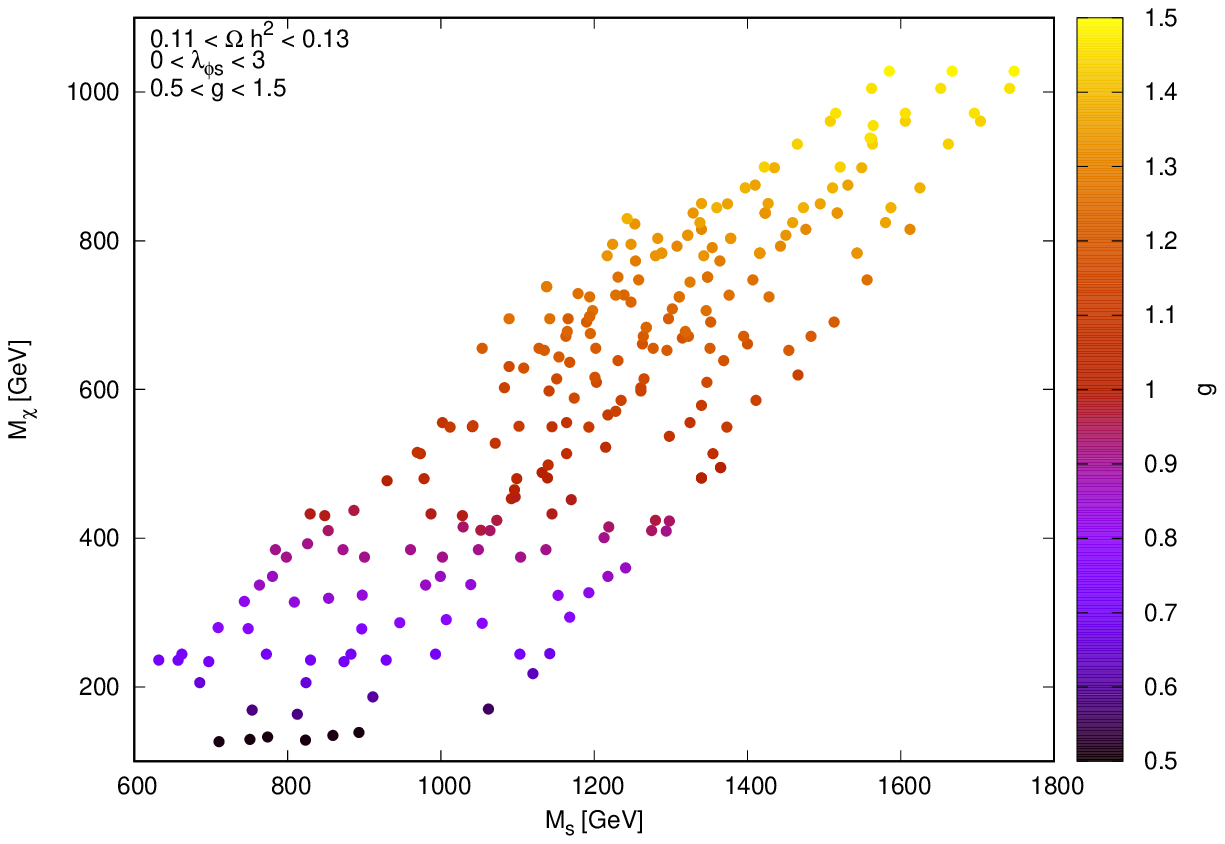,width=7.5cm}\hspace{0cm}\epsfig{figure=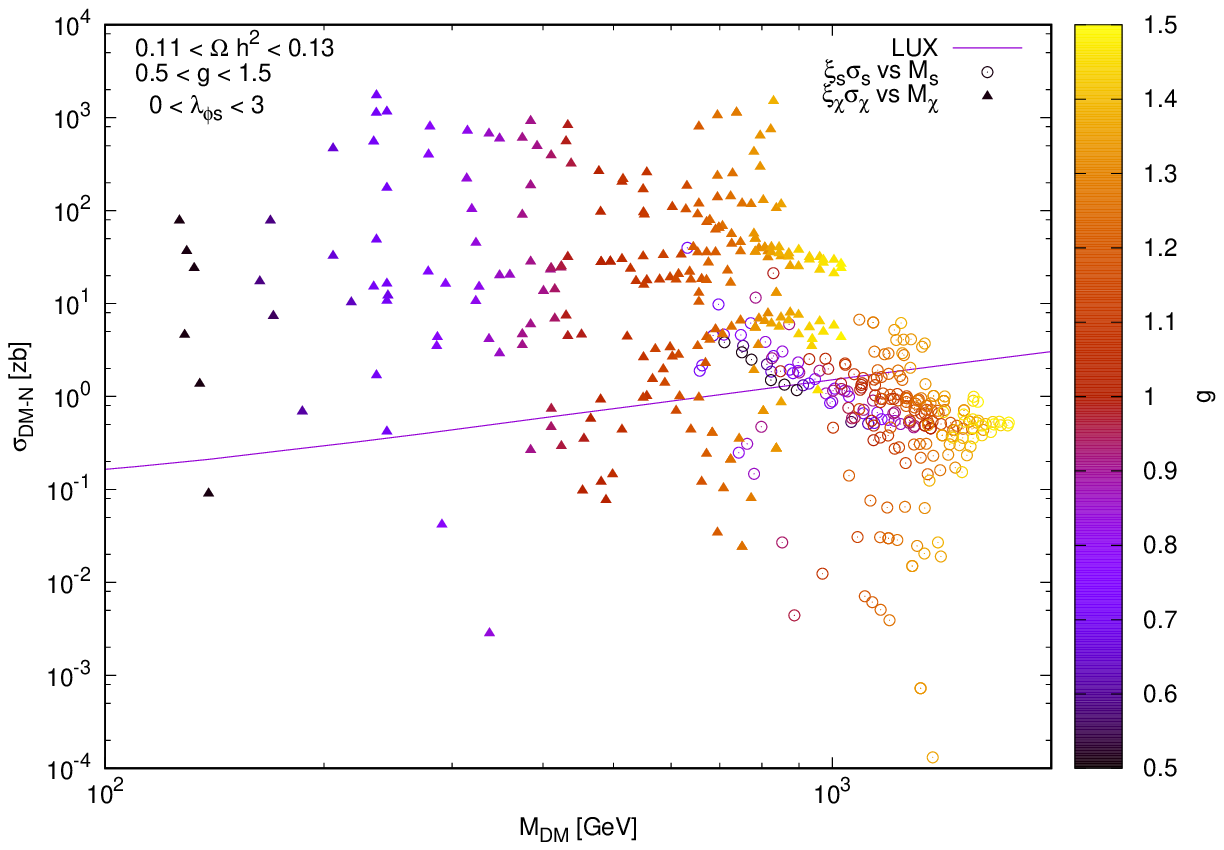,width=7.5cm}}
\centerline{\hspace{0cm}\epsfig{figure=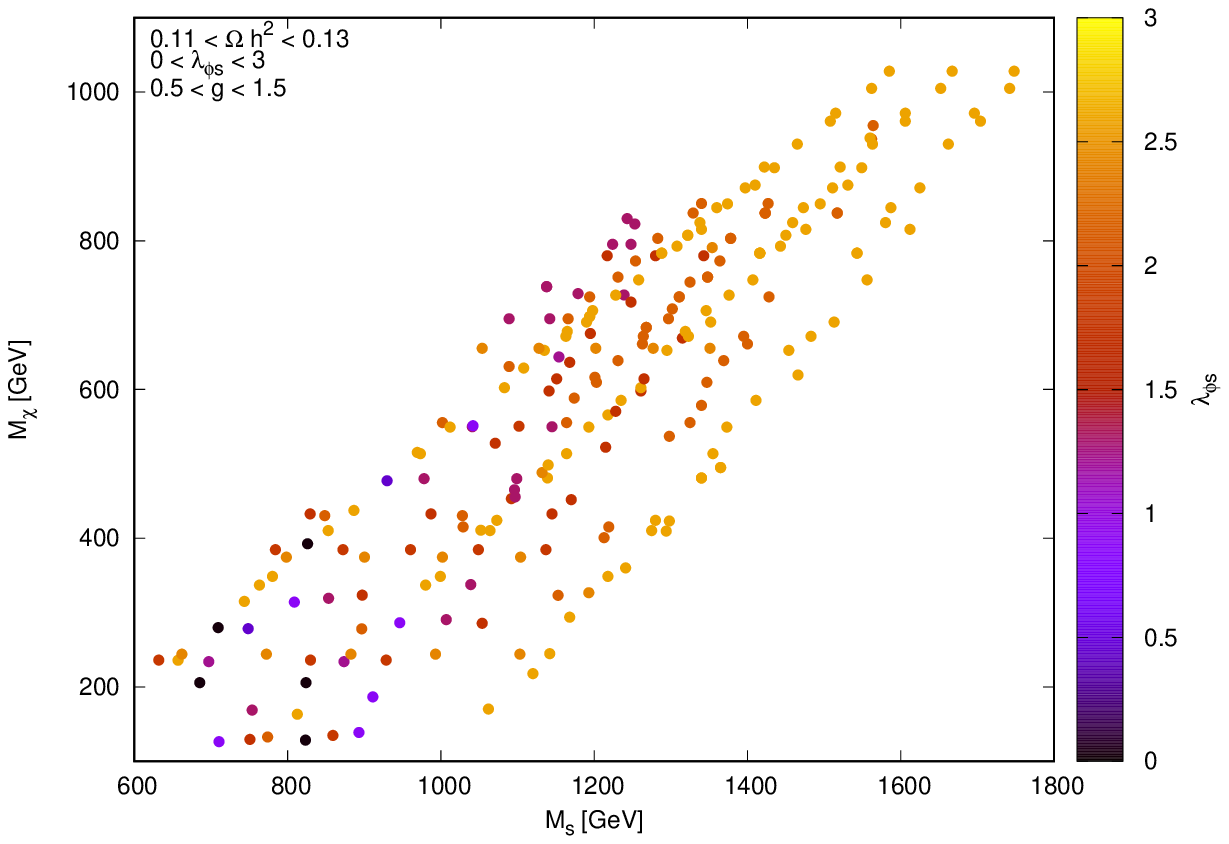,width=7.5cm}\hspace{0cm}\epsfig{figure=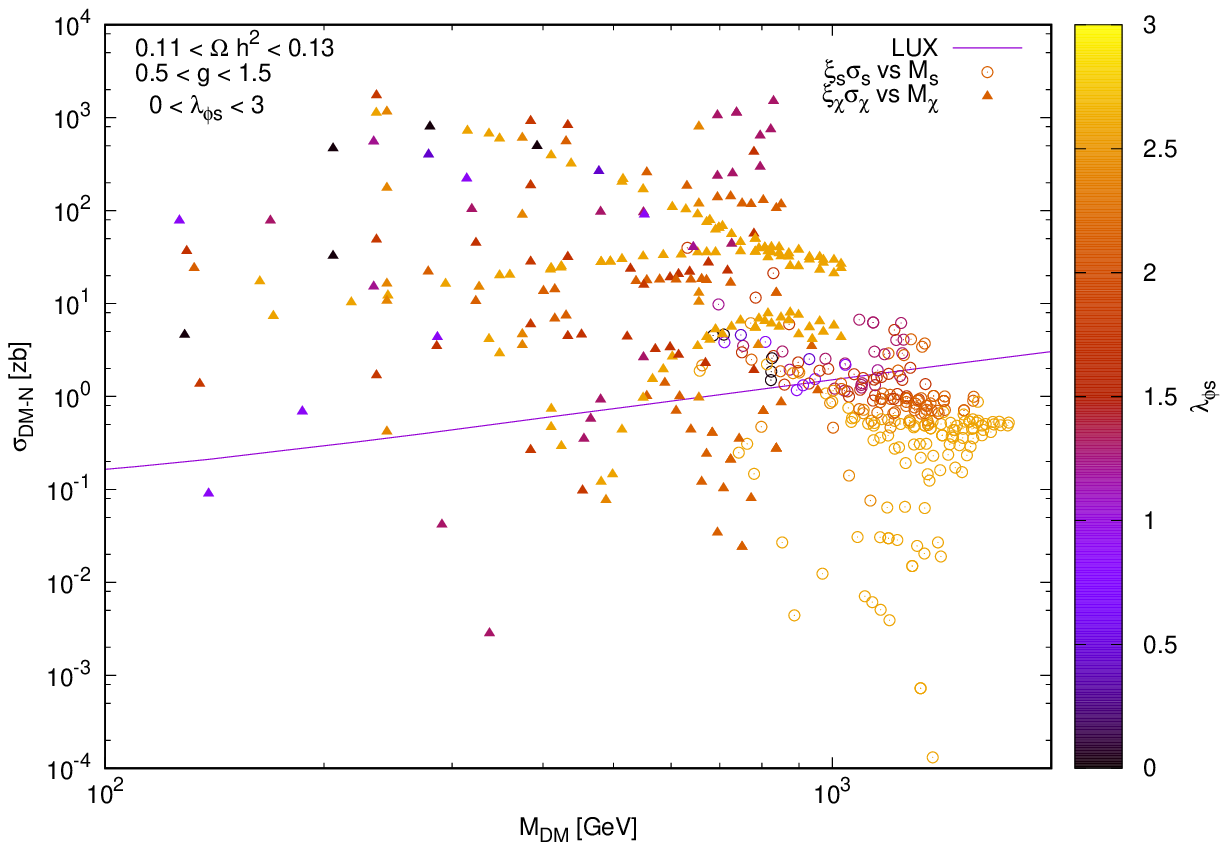,width=7.5cm}}
\centerline{\hspace{0cm}\epsfig{figure=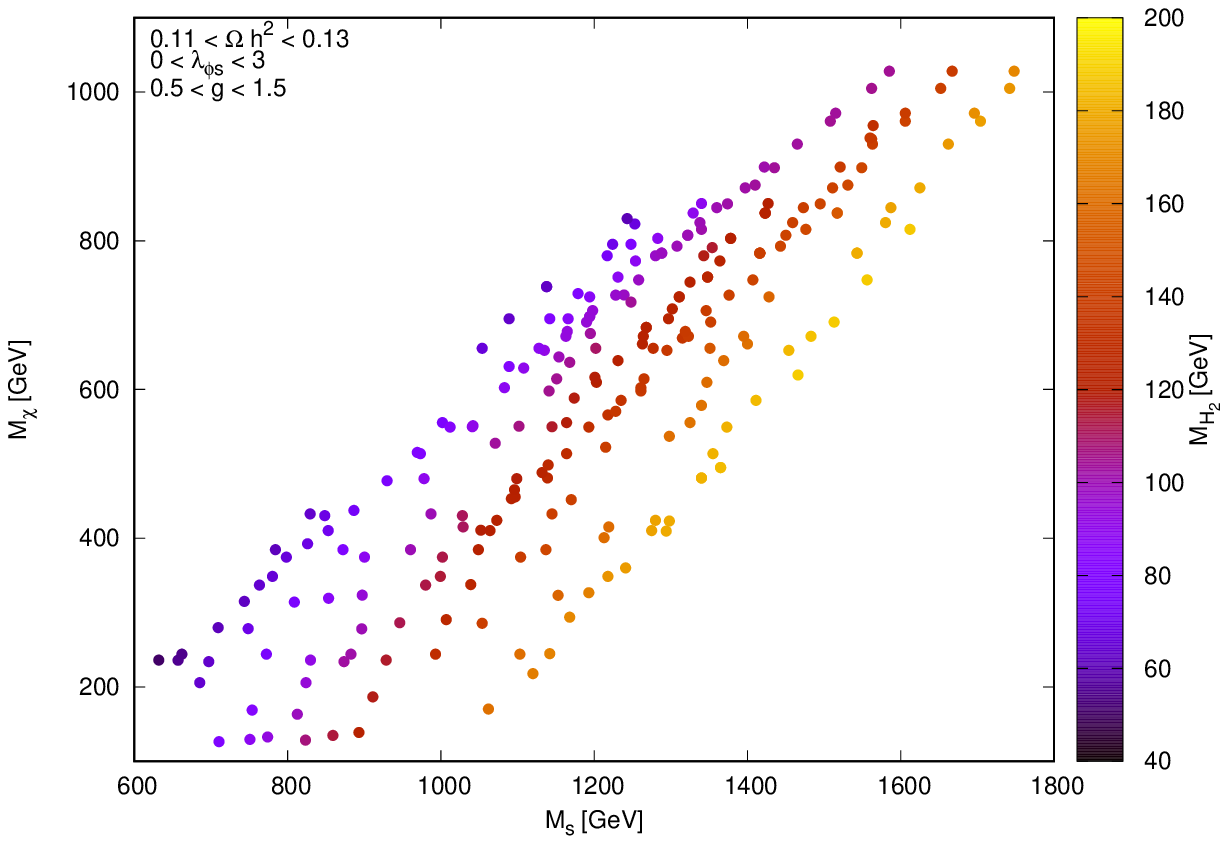,width=7.5cm}\hspace{0cm}\epsfig{figure=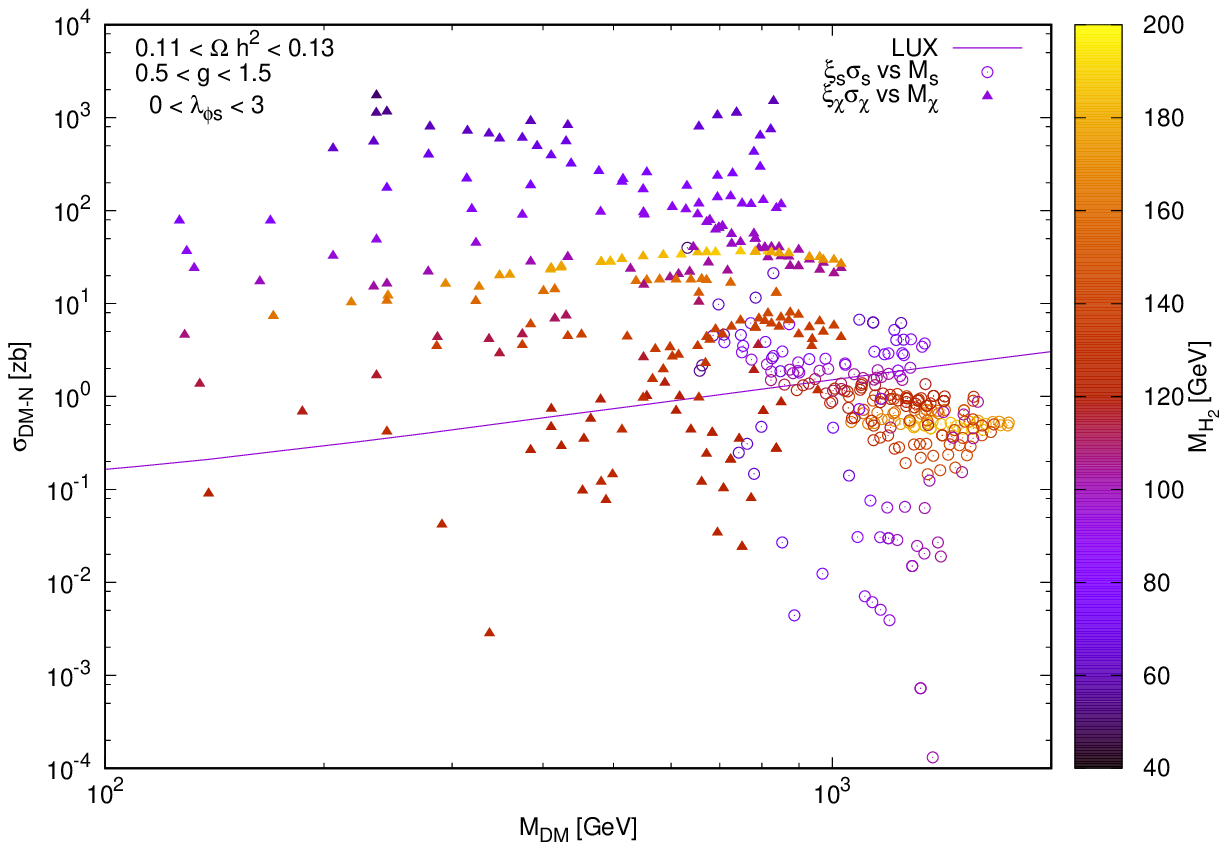,width=7.5cm}}
\centerline{\vspace{-0.7cm}}
\caption{(Left) Scatter points depict ranges of parameters space of the model in $M_s$ and $M_{\chi}$ plane for different parameters of the model which are consistent with observed relic density by Planck collaboration \cite{Planck}. (Right) depicts rescaled DM-Nucleon cross section as function of DM mass for different values of other model parameters.}\label{constrained Relic Density and Direct Detection}
\end{center}
\end{figure}

Notice that in above analysis, we separately suppose $\xi_S=1$ and $\xi_{\chi}=1$ in Fig.~\ref{Direct}-a and Fig.~\ref{Direct}-b. In next step, we display combine analysis, direct detection and relic density in Fig.~\ref{constrained Relic Density and Direct Detection}. In order to study the effect of the direct detection experiment on the model,  rescaled DM-Nucleon cross section $\xi_{\chi}\sigma_{\chi}$ and $\xi_S\sigma_S$  should be considered. Scatter points in Fig.~\ref{constrained Relic Density and Direct Detection} (Left) show allowed region in parameters space of the model in $M_s$ and $M_{\chi}$ plane for different parameters of the model which are consistent with observed relic density by Planck collaboration \cite{Planck}. In this figures, it is supposed  $ 0.11 < \Omega h^{2} < 0.13 $ for allowed range of relic density and  also $ 0 < \lambda_{\phi s} < 3 $, and $ 0.5 < g < 1.5 $. Right figures depict rescaled DM-Nucleon cross section verses DM mass  for different values of other model parameters. The miles line determines upper limit of LUX experiments for direct detection of DM.

\subsection{Indirect Detection}
\begin{figure}[!htb]
\centerline{\hspace{0cm}\epsfig{figure=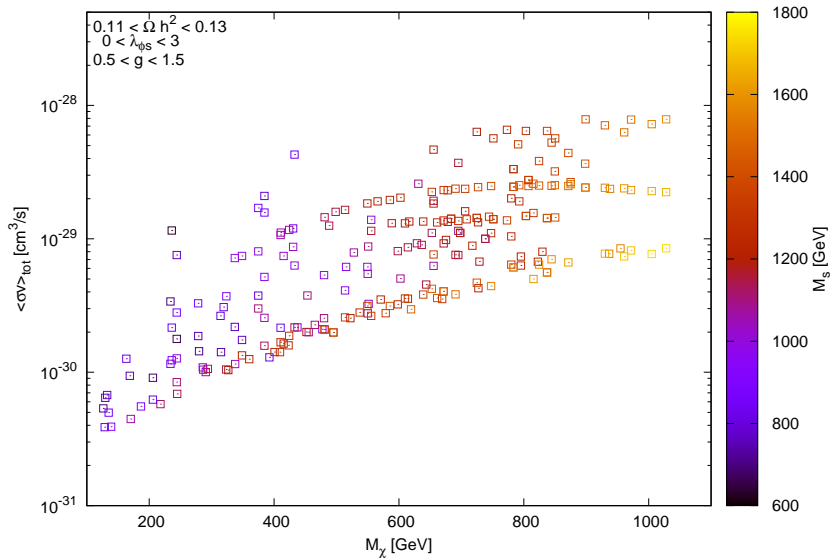,width=11cm}}
\caption{Velocity-averaged annihilation cross
section of DM as two component DM mass}\label{DM-Annihilation}
\end{figure}

The indirect detection of DM annihilation and decay using observations of photons, charged cosmic rays, and neutrinos offers a promising means of identifying nature of this part of Universe.  There are currently intensive international efforts to detect these astroparticles as signature of DM particles. In the freeze-out scenario, the pair annihilation rate of a  thermal relic DM particle is directly linked to the today relic abundance. 
Based on the measured abundance of DM, a particle which  constitutes
all of the DM will have a total pair annihilation cross section of $ <\sigma v> \, \sim O(10^{-26}) \, cm^{3}/s $\cite{indirect}. This value is often used as a benchmark and is referred to as the thermal relic cross section. We have calculated the velocity-averaged annihilation cross
section of DM for $ 0.11 < \Omega h^{2} < 0.13 $, $ 0 < \lambda_{\phi s} < 3 $, and $ 0.5 < g < 1.5 $ by using micrOMEGAs  package \cite{Belanger:2014vza}.  Our result is shown in Figure \ref{DM-Annihilation}. As it is seen, the results can not saturate particle fluxes detected in aforementioned indirect detection experiments limits. This means astoparticle fluxes which are coming from the galactic center, should have other astrophysical origins.

\subsection{Self-Interaction}
The self interaction of DM can potentially be probed by studying the offset between the DM halo and the stars of a galaxy moving through a region of large DM density. The first evidence for DM self interactions has been reported \cite{Abel 3827} based on observations of four elliptical galaxies in the inner $10~\rm kpc$ core of galaxy cluster Abell 3827.  An updated
work \cite{Kaplinghat:2015aga} has considered a set of twelve galaxies and six clusters in order to cover different scales. Including the core sizes from dwarf to cluster (varying from 0.5 to 50
kpc), the aforementioned cross section is parametrized
as 
\begin{eqnarray}
\sigma^{eff}_{self}/m_{DM}\sim 0.1-2~cm^2g^{-1}
\end{eqnarray}
where the effective self-interacting cross section is defined by
$\sigma^{eff}_{self}/{m_{DM}}=$$\xi^2_{\chi, S}\frac{\sigma_{self}}{m_{DM}}$ and $\xi_{\chi,S}$ is the fraction of one of two DM component. In particle physics
units, this corresponds to $\sigma^{eff}_{self}/m_{DM}\sim( 0.43-8.72)\times 10^3~\rm GeV^{-3}$.

\begin{figure}[!htb]
\begin{center}
\centerline{\hspace{1cm}\epsfig{figure=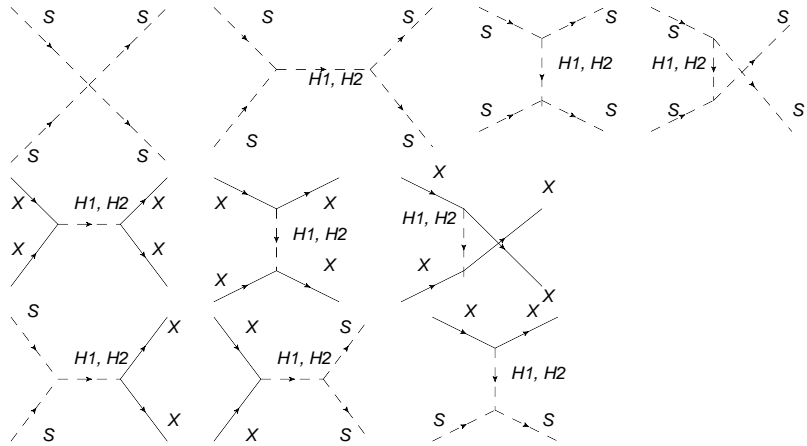,width=10.5cm}}
\centerline{\vspace{-0.8cm}}
\end{center}
\caption{The Feynman diagrams for scalar and fermion DM
self-interactions.}\label{self}
\end{figure}

In next step, we consider the DM
self-interacting cross section for scalar $S$ and fermion $\chi$
DM. The DM self interactions include processes:  $SS\longrightarrow SS$,
$\chi\chi\longrightarrow \chi\chi$, $SS\longrightarrow \chi\chi$,
$\chi\chi\longrightarrow SS$ and  $S\chi\longrightarrow S\chi$.
Fig.~\ref{self} shows Feyman diagrams for DM self interactions.

\begin{figure}[!htb]
    \begin{center}
        \centerline{\hspace{0cm}\epsfig{figure=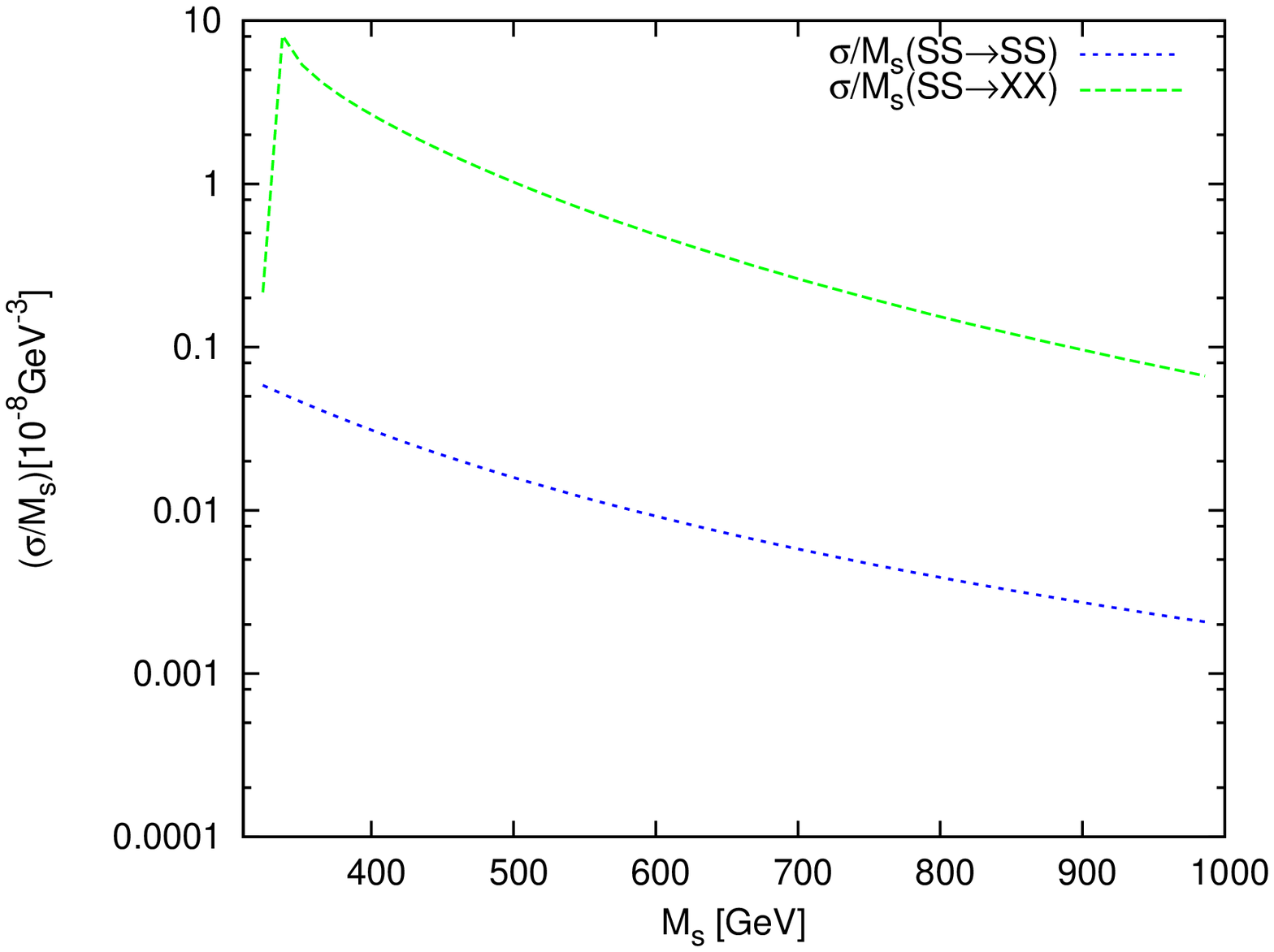,width=7.5cm}\hspace{0.2cm}\epsfig{figure=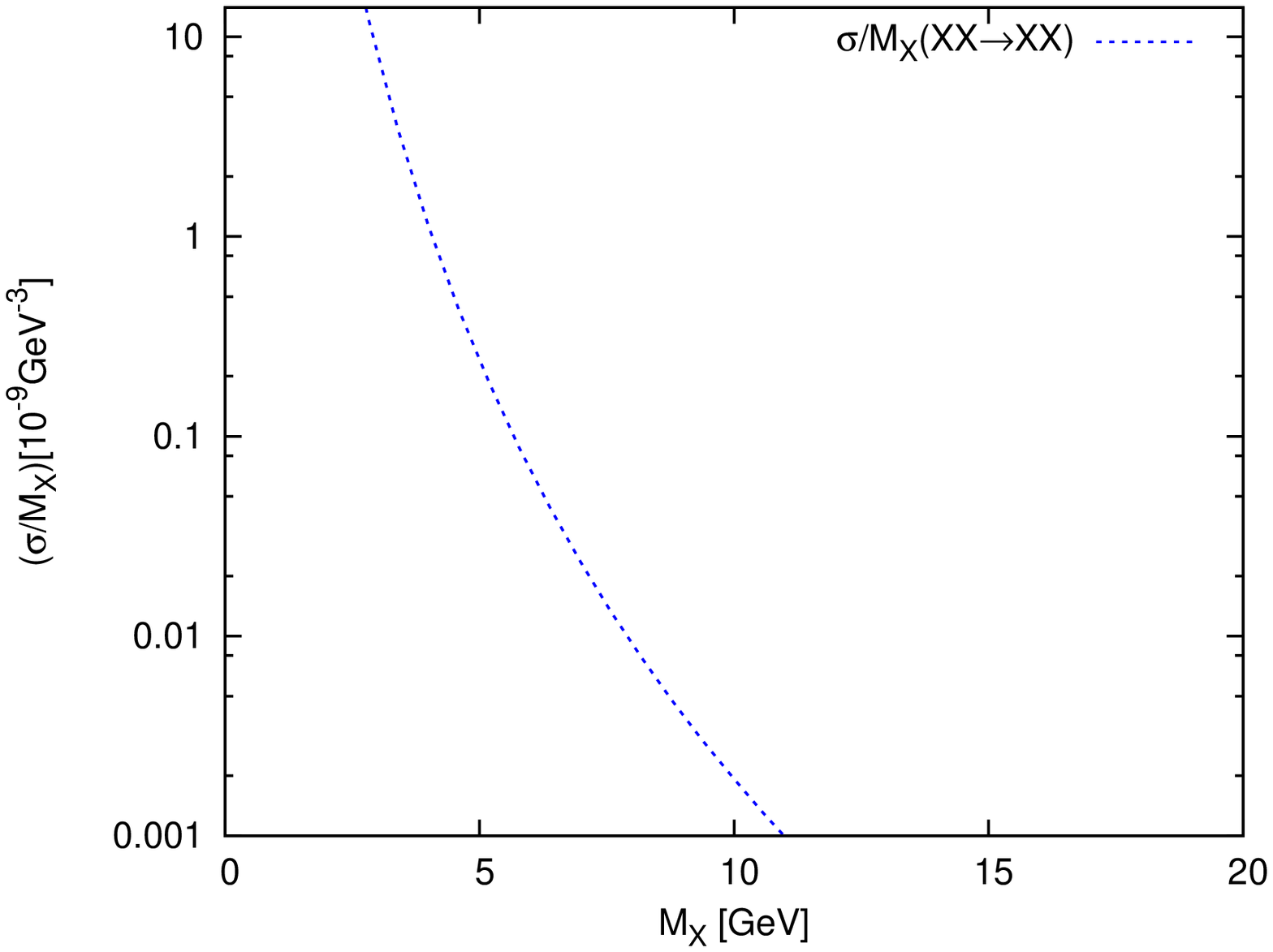,width=7.5cm}}
        \centerline{\vspace{-1.5cm}\hspace{1cm}(a)\hspace{6cm}(b)}
        \centerline{\vspace{-0.0cm}}
    \end{center}
    \caption{a) The scalar self-interaction cross section as a function of scalar DM mass. We set  $M_{\chi}=100~\rm GeV$ and $g=0.5$ $\lambda_{\phi s}=2$ and $\lambda_s=2$. b) The fermionic self-interaction cross section as a function of fermion DM mass. Input parameters are similar to (a) except  $M_{s}=500~\rm GeV$.}\label{curveself}
\end{figure}

The main contributions to $\sigma/M_{s}$ for scalar annihilation (processes $SS\longrightarrow SS$\cite{scalar-self},
and $SS\longrightarrow \chi\chi$) are given in appendix. For process $SS\longrightarrow SS$\cite{scalar-self},  $\sigma/M_{s}$ is proportional to $1/M^3_s$ and after imposing constraint $M_s>310~\rm GeV$, we find that this situation does not saturate  upper bound on the self-interaction cross section. Indeed, to obtain reasonably strong scalar DM self-interaction, mass of scalar must
be very small, $M_ s<1~\rm GeV$. Since in non relativistic regime $s\sim4M_s^2$, $\sigma(SS\rightarrow \chi\chi)/M_s$ will be larger than $\sigma(SS\rightarrow SS)/M_s$, This feature depicts in Fig.\ref{curveself}-a. As it is seen in this figure, self-interaction for scalar DM is very smaller than upper bound. However it is possible to achieve upper
bound on the self-interaction cross section for scalar DM if we consider self-interaction in the vicinity of resonance $M_s\simeq M_{H_2}/2$. Note that according to Eq.\ref{11}, the mass of scalar DM can not be equal to half of SM Higgs mass. For resonance regime ($M_s\simeq M_{H_2}/2$) , the s-channel $H_2$ exchange diagram in Fig.~\ref{self} dominates and scalar DM self-interaction may exceed experimental bound. Achieving the observed scalar DM self-interaction cross section requires that $M_s$ be severely tuned such that $|M_s-M_{H_2} /2|< 1~ \rm MeV$ (While $M_s>310 ~\rm GeV$). 
However, since the main contribution of observed relic density was obtained from fermionic DM and scalar DM has small contribution to relic density, we expect that this process is very rare in the center of Milky Way galaxy.     

In the following, we consider self-interaction for the case of Dirac fermionic DM which includes processes $\chi\chi\longrightarrow \chi\chi$ and $\chi\chi\longrightarrow SS$. The main Feynman diagram which contribute to  aforementioned  process are s channel for $\chi\chi\longrightarrow \chi\chi$ in Fig.~\ref{self} and $\chi\chi\longrightarrow SS$.  The cross section of these processes are presented in appendix. 
 
For the process $\chi\chi\longrightarrow SS$ in non-relativistic limit $s < 4M^2_s$ and so this process is forbidden. For processes $\chi\chi\longrightarrow \chi\chi$, since in non relativistic regime $s\simeq4M_{\chi}^2$, self-interaction of fermionic DM is much smaller than experimental bound (it has been shown in Fig.~\ref{curveself}-b). It also turns out that to vitalize reasonably strong fermionic DM self-interaction (similar to scalar DM), we should consider self-interaction in the near resonance $M_{\chi}\simeq M_{H_2}/2$ or $M_{H_1}/2$. Notice that for fermionic DM fine tuning should be stronger than scalar DM due to smaller self-interacting cross section for fermionic DM.  

\begin{figure}[!htb]
     \begin{center}
         \centerline{\hspace{0cm}\epsfig{figure=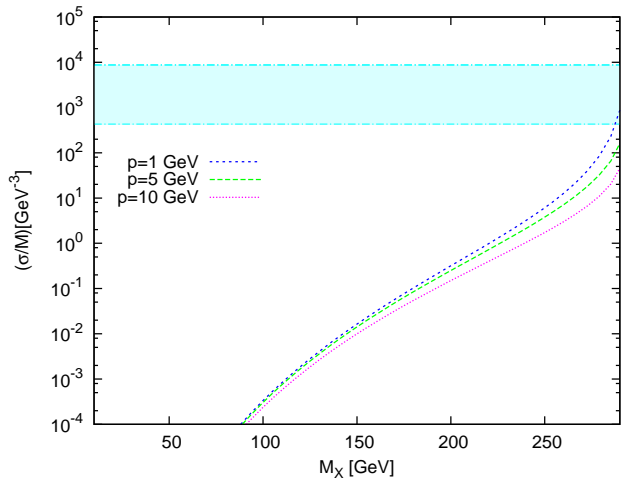,width=8.5cm}} \centerline{\vspace{-1.5cm}}
     \end{center}
     \caption{The scalar-fermion self-interaction cross section as a function of fermion DM mass. We set  $M_{s}=500~\rm GeV$ and $g=0.5$ $\lambda_{\phi s}=0.5$, $\lambda_s=0.5$ and different values for momentum of initial fermionic DM. The shadowed panel indicates allowed range of experimental measurements for DM  self-interaction.}\label{self-Sf}
 \end{figure}

In continue, we also calculate the DM self interaction cross-sections for processes $S\chi\longrightarrow S\chi$ in non-relativistic limit. The cross sections is given in appendix.  
Given the fact that the main contribution of observed relic density was obtained from fermionic DM and contribution of the scalar DM is less than 1 percent of total relic density, occurrence of this process is very rare.  To estimate the magnitude of $\sigma_{S\chi\longrightarrow S\chi}$ in non-relativistic limit, we suppose  $s\simeq(M_{\chi}+M_s)^2$ and also consider $\overline{M}=\frac{(M_s+M_{\chi})}{2}$. Note that this process does not affect the relic density of DM. In Fig.~\ref{self-Sf}, we depict  the contribution of $S\chi\longrightarrow S\chi$ versus  $\overline{M}$ for several values of initial momentum of fermionic DM. As it is seen, the specified process does not contribute to this cosmological constraint. In this estimation, we did not consider the difference in the fraction of two DM component.

\subsection{Invisible Higgs decay}
\begin{figure}
\begin{center}
\centerline{\hspace{0cm}\epsfig{figure=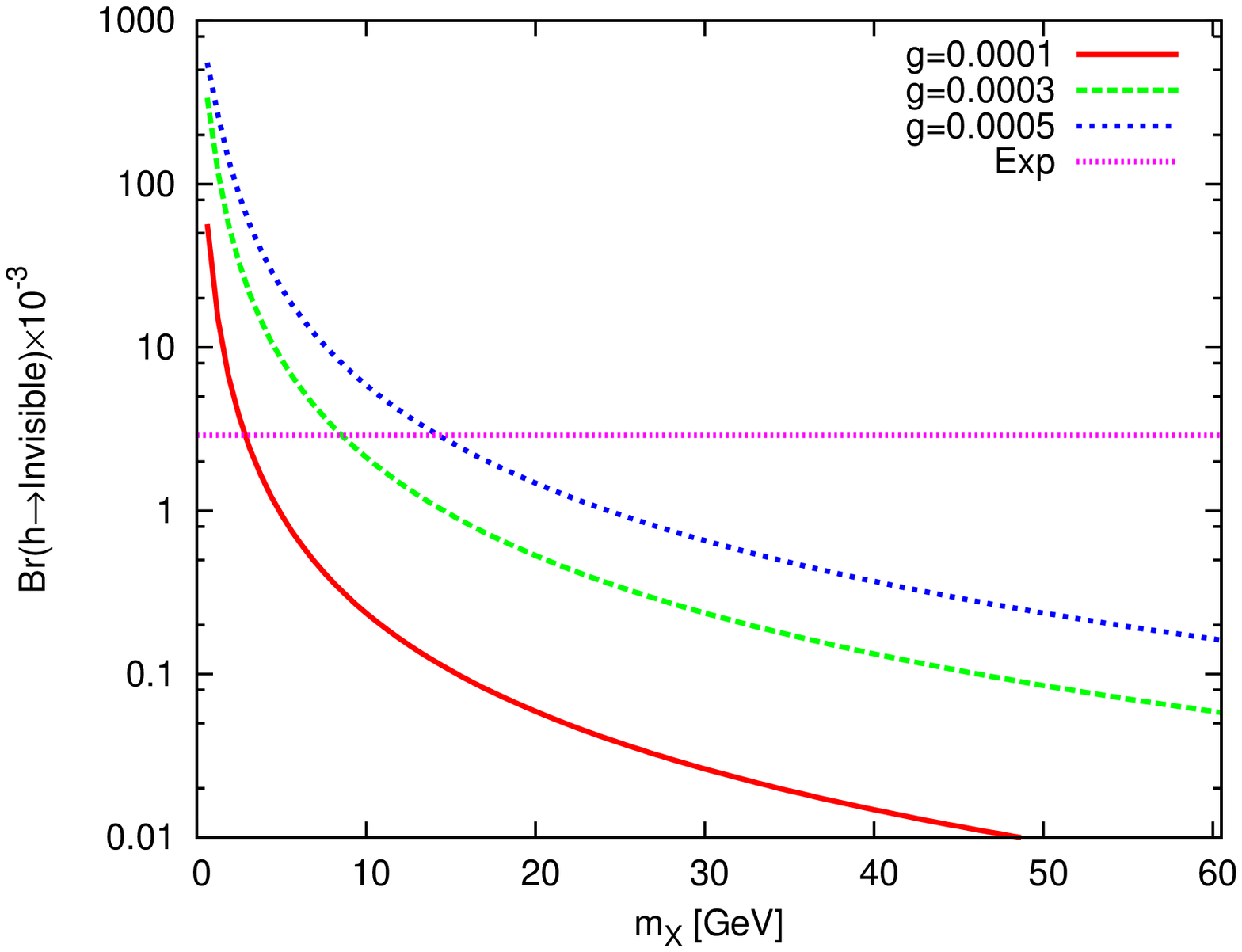,width=7.5cm}\hspace{0.3cm}\epsfig{figure=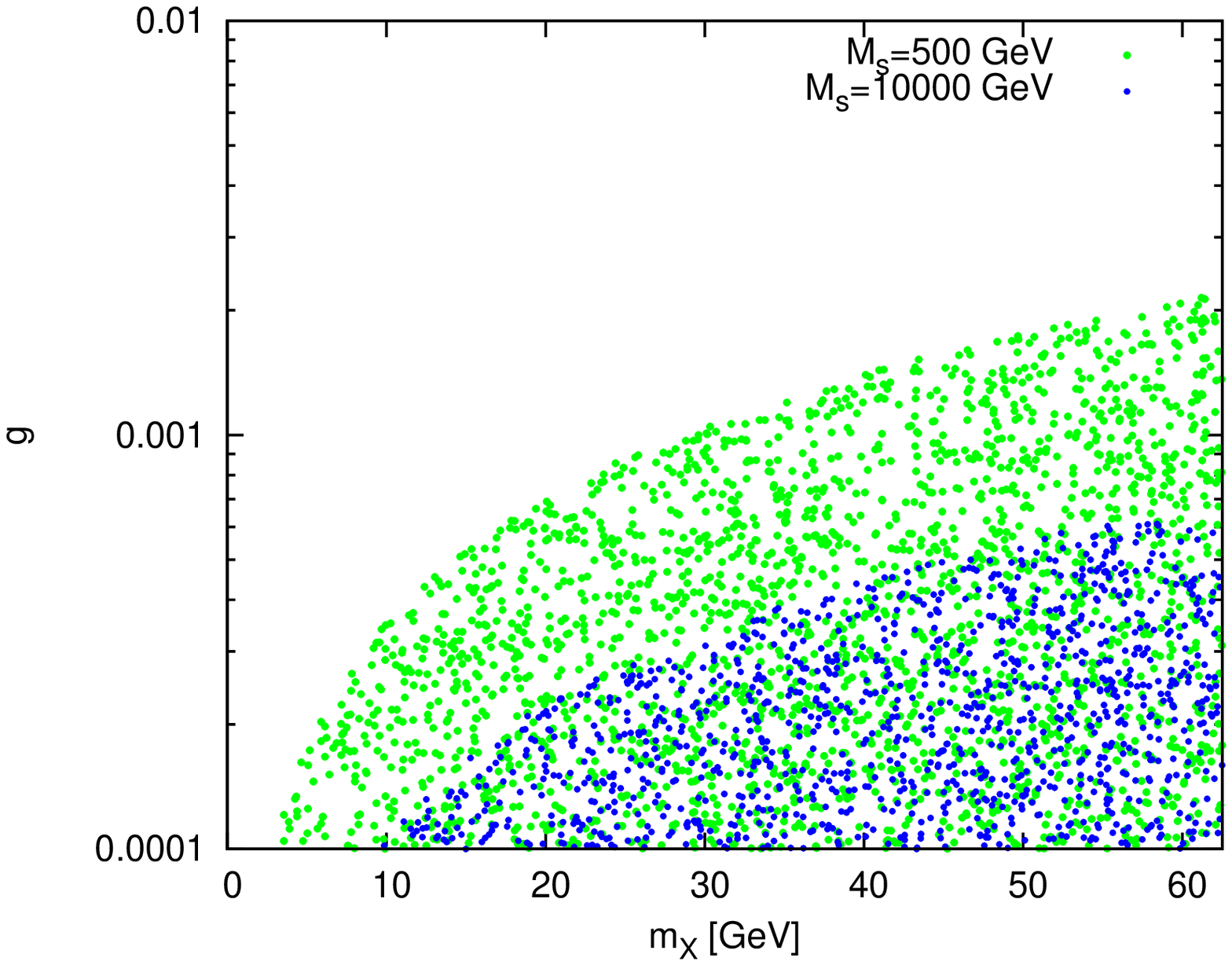,width=7.5cm}}     \centerline{\vspace{-1.5cm}\hspace{0.5cm}(a)\hspace{6cm}(b)}
\centerline{\vspace{-0.0cm}}
\end{center}
\caption{a) $Br(H_1\rightarrow \rm Invisible)$ as function of fermionic DM mass for different values of $g$ coupling and $M_s=500~\rm GeV$. b) Scater points depict ranges of parameters space in mass of fermionic DM and  $g$ for different values of $M_s$ which are consistent with experimental measurements of $Br(H_1\rightarrow \rm Invisible)$.}\label{INV}
\end{figure}

\begin{figure}
\begin{center}
\centerline{\hspace{0cm}\epsfig{figure=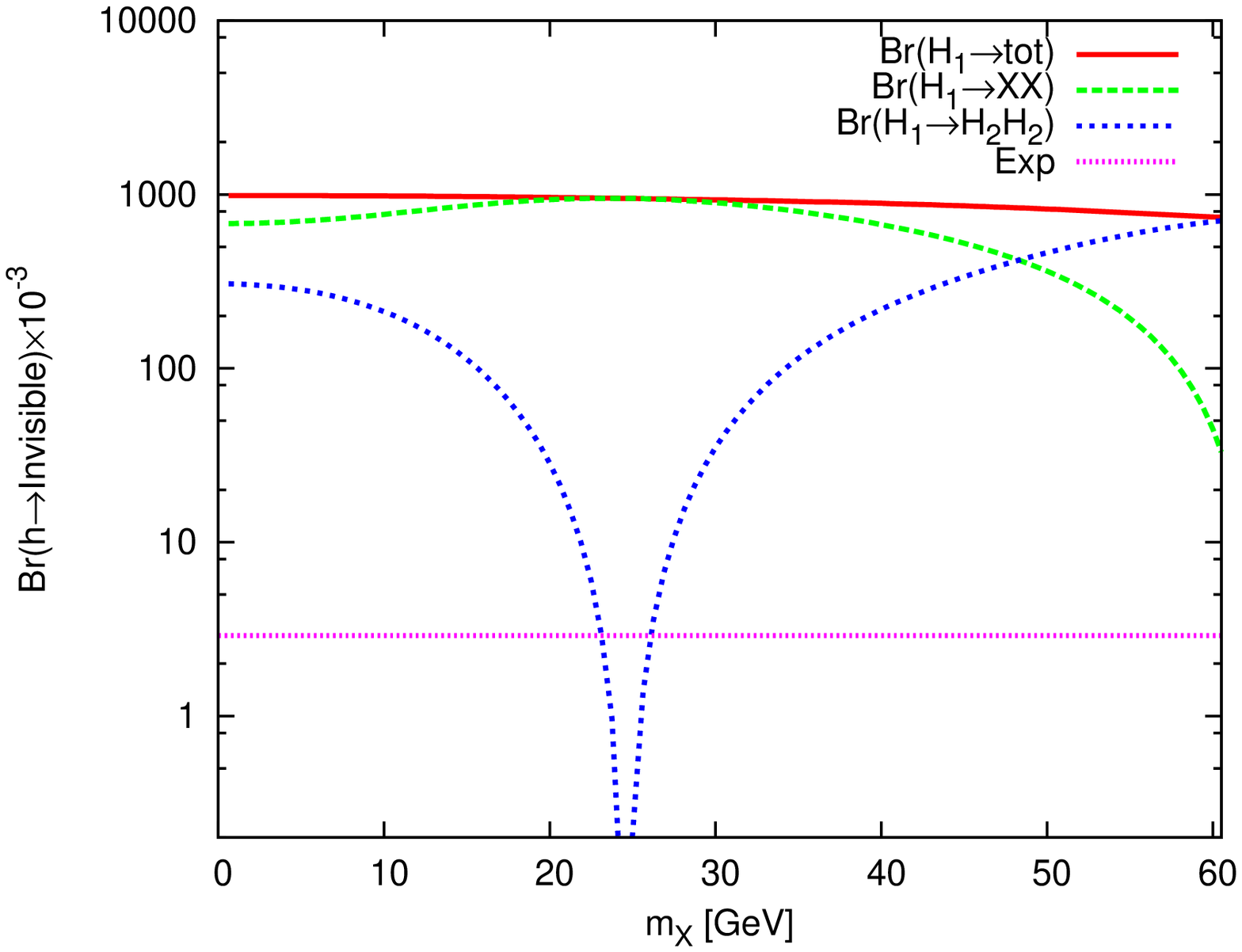,width=7.5cm}\hspace{0.2cm}\epsfig{figure=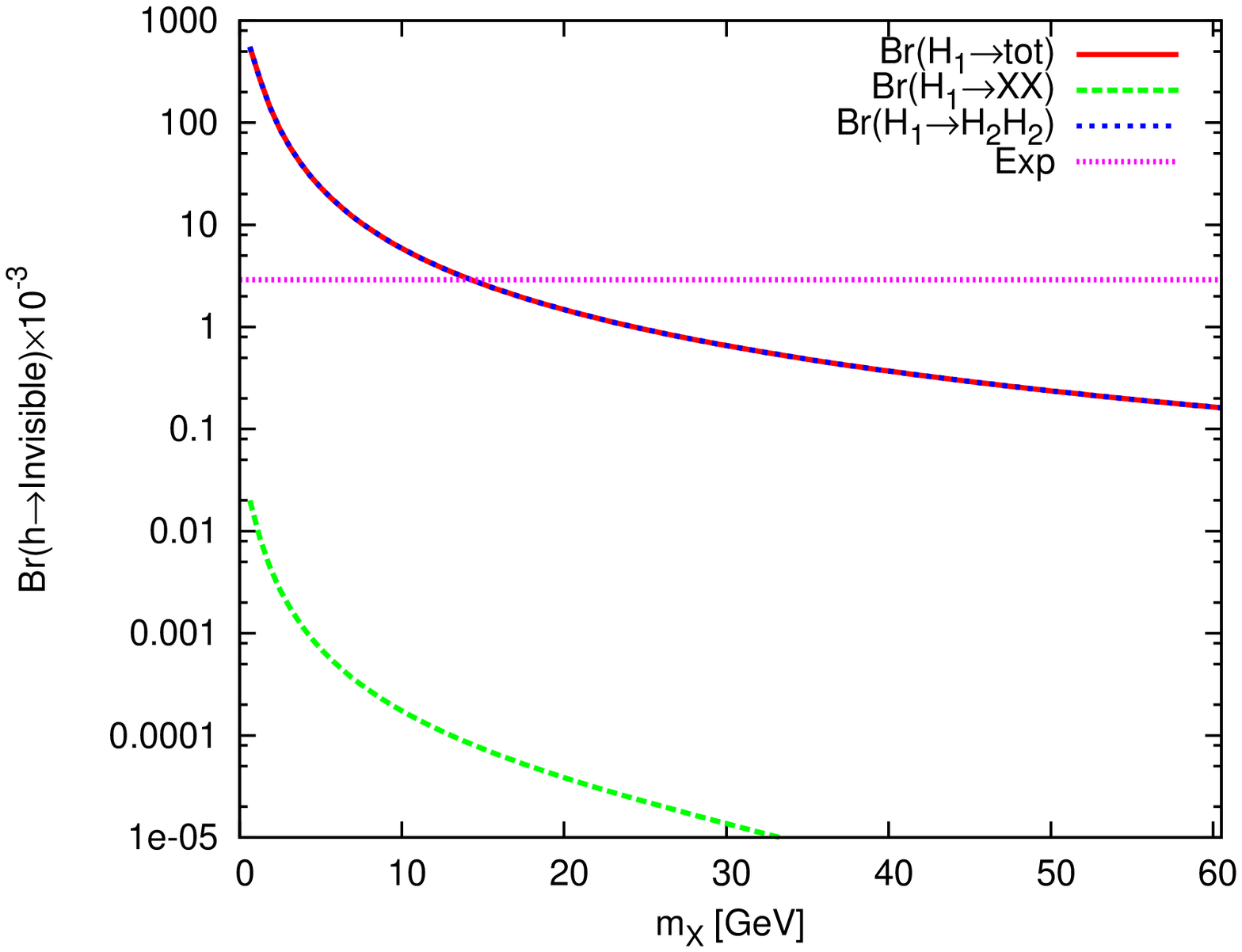,width=7.5cm}}     \centerline{\vspace{-1.5cm}\hspace{0.5cm}(a)\hspace{6cm}(b)}
\centerline{\vspace{-0.0cm}}
\end{center}
\caption{a) (b) Different contribution of $Br(H_1\rightarrow \rm Invisible)$ as function of fermionic DM mass for $g=0.1$ (g=0.0005) and  $M_s=500~\rm GeV$.}\label{INVC}
\end{figure}

The observed
Higgs boson at 125 GeV, might decay to a component of DM  which does not interact with the detector. Therefore it opens a window for exploring possible DM-Higgs boson coupling. 
Notice that invisible Higgs boson decays are only sensitive to DM
coupling in region of parameters space which are kinematically allowed.
 Here, we suppose $H_1$ is the SM Higgs boson as a result, if scalon, scalar and fermionic DM are lighter than SM Higgs boson, they can contribute to the
 invisible decay mode of Higgs boson with branching ratio:
\begin{align}
Br(H_1\rightarrow \rm Invisible) =\frac{\Gamma(H_1\rightarrow
\chi\chi)+\Gamma(H_1\rightarrow
SS)+\Gamma(H_1\rightarrow
H_2H_2)}{\Gamma(h)_{SM}+\Gamma(H_1\rightarrow SS)+\Gamma(H_1\rightarrow
\chi\chi)+\Gamma(H_1\rightarrow
H_2H_2)},\label{decayinv1}
\end{align}
where $\Gamma(h)_{SM}=4.15 ~ \rm [MeV]$ is total width of Higgs
boson \cite{SMinv}. The decay rates for $H_1\rightarrow \chi\chi$, $H_1\rightarrow SS$ and $H_1\rightarrow H_2H_2$ have been presented in appendix.
Branching ratio of invisible Higgs mode has been constrained by various groups using the latest data from LHC \cite{CMS:Inv,ATLAS:Inv,InvHiggs}. 
ATLAS Collaboration has reported a search of the SM
Higgs boson decay in its invisible decay mode and obtaining  an
upper limit of $75\%$, at a mass of 125.5 GeV\cite{InvHiggs}.
In the SM, the main process which contribute to invisible decay of the Higgs boson is $h\rightarrow ZZ^*\rightarrow 4\nu$, but $Br(h\rightarrow ZZ^*\rightarrow 4\nu)=1.2\times10^{-3}$ \cite{Heinemeyer:2013tqa} is below the sensitivity of the ATLAS collaboration analysis. According to Eq.~\ref{11},
$M_s>310 ~\rm GeV$ and so SM Higgs boson $H_1$ can not decay
to scalar DM. In Fig~.\ref{INV}-a, we display $Br(H_1\rightarrow \rm Invisible)$ as a function of fermionic DM mass for different values of  $g$ coupling. In this figure, we suppose $M_{\chi}<M_{H_1}/2$ and assign other parameters such that $M_{H_2}<M_{H_1}/2$. By using ATLAS upper limit for invisible Higgs decay, we display allowed range of parameters space in Fig~.\ref{INV}-b in our model.  Note that the main contribution to $Br(H_1\rightarrow \rm Invisible)$ in the portion of parameters space which is consistent with experimental limits arises from $\Gamma(H_1\rightarrow H_2H_2)$. This feature has been shown in Fig.~\ref{INVC}.  This figure separately depicts contribution of $Br(H_1\rightarrow \rm Invisible)$ as a function of the fermionic DM mass 
for $Br(H_1\rightarrow \chi\chi)$, $Br(H_1\rightarrow  H_2H_2)$ and $Br(H_1\rightarrow \rm total)$. Comparing Fig.~\ref{INVC}-a and b implies for small values of $g$ which is consistent with experimental limits, the main contribution of $Br(H_1\rightarrow \rm Invisible)$ are coming from $Br(H_1\rightarrow  H_2H_2)$.  In our model,  $M_{H_2}$ generally depends on $g$, $M_{\chi}$ and $M_s$.  
Since $\Gamma(H_1\rightarrow H_2H_2)$ depends on $M_{H_2}$, in allowed region of parameters space, we expect that the branching ratio of invisible Higgs decay also depends on $M_s$. In Fig.~\ref{INV}-b, we have shown for larger values of $M_s$, allowed area shrinks in $g_s$ and $M_{\chi}$ plane.   

\begin{figure}
\centerline{\hspace{0cm}\epsfig{figure=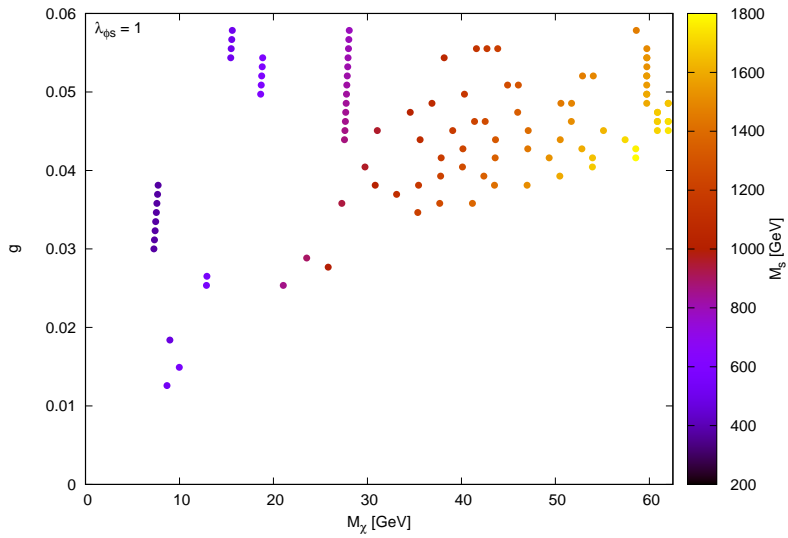,width=10.5cm}}
\caption{Scater points depict ranges of parameters space in mass of fermionic DM and $g$ coupling for different values of $M_s$ which are consistent with observed relic density.}\label{plot18}
\end{figure}

In Fig.~\ref{plot18}, ranges of parameters space in mass of fermionic DM and $g$ coupling which are consistent with observed relic density have been shown. Comparing Figures \ref{plot18} and \ref{INV}-b shows that the allowed region for invisible Higgs decay and the DM relic density  does not overlap with each other. Since the most contribution of DM relic density arises from fermionic DM, for small value of $g$ coupling, the annihilation of DM to SM particles will be suppressed.
This means for portion of parameters space which is consistent with invisible Higgs decay, the relic density exceed the value of Planck measurement. Therefore, in order to evade invisible Higgs constraints, one should assume that fermionic DM mass is larger than $ \frac{M_{H_{1}}}{2} $.
\section{Concluding remarks}
Motivated by DM and hierarchy problems, we presented a scale-invariant extension of SM. In order to have a scale invariant version of the SM with scalar DM,  at least two more scalars must be added to the theory. Moreover, in the absence of additional fermionic fields, the model has a small number of independent parameters which complicates the satisfying all theoretical and phenomenological constraints. Given these conditions, we added a scalon field $\phi$, a scalar field $S$ and a fermionic field $\chi$ as two-component DM to SM. To summarize, the main novelty of this model with respect to other two-component DM models, is a much lower number of independent parameters due to scale invariant conditions.

In this analysis, relic density of two component DM was computed. We have shown that the most part of contribution of DM relic density  arises from fermionic DM.
We have discussed the allowed regions in parameter space of our model consistent with the observed relic density.

We have also taken into account the constraints of indirect detection and direct detection of DM. 
In order to constrain the parameter space of our
model, we also checked the limits from self interaction of DM.  It is shown that the former analysis can not put constraint on the model in large portion of parameters space. Only in the vicinity of the resonances in $M_s\simeq M_{H_2}/2$ for scalar DM and $M_{\chi}\simeq M_{H_2}/2$ or $M_{H_1}/2$  for fermionic DM, self-interaction scenario constraints the model.

Finally, we probed the limits from the invisible decay width of the Higgs. We have found that the viable regions in parameter space are in agreement with upper limit on the invisible Higgs decay branching ratio. We compared the consistent region in
parameter space for invisible Higgs decay with the relic density of the fermionic DM and show that in order to satisfy invisible Higgs constraints, fermionic DM mass should be larger than $ M_{H_{1}}/{2}$.

\section*{Acknowledgement}
The authors would like to thank S. Paktinat for careful reading of the manuscript and the useful remarks.

\section*{Appendix:  DM self-interaction cross sections and Decay Rates}

In this appendix, we summarize the formula of the self-interacting cross-sections for two components of DM and decay rates of two scalars Higgs.

The main contribution to $\sigma/M_{s}$ for scalar annihilation (processes $SS\longrightarrow SS$\cite{scalar-self},
and $SS\longrightarrow \chi\chi$) in the non-relativistic limit are given by:
\begin{align}
\sigma(SS\rightarrow SS)/M_s&=\frac{1}{64\pi M^3_s}|\lambda_s 
+ \frac{2\lambda_{\phi s}M_{\chi}/g}{\sqrt{1+(\nu_1g/M_{\chi})^2}}\frac{1}{s-M_{H_2}^2+iM_{H_2}\Gamma_{H_2}}\nonumber\\&- \frac{2\lambda_{\phi s}\nu_1}{\sqrt{1+(\nu_1g/M_{\chi})^2}}\frac{1}{s-M_{H_1}^2+iM_{H_1}\Gamma_{H_1}}|^2,
\label{scalar}
\end{align}
\begin{align}
\sigma(SS\rightarrow \chi\chi)/M_s&=\frac{1}{32\pi M_s }(1-\frac{4M_{s}^2}{s})^{-1/2}(1-\frac{4M_{\chi}^2}{s})^{3/2}\nonumber\\
&\times|\frac{2\lambda_{\phi s}M_{\chi}}{1+(\nu_1g/M_{\chi})^2}\frac{1}{s-M_{H_2}^2+iM_{H_2}\Gamma_{H_2}}\nonumber\\&+ \frac{2\lambda_{\phi s}\nu^2_1g^2/M_{\chi}}{1+(\nu_1g/M_{\chi})^2}\frac{1}{s-M_{H_1}^2+iM_{H_1}\Gamma_{H_1}}|^2,
\label{ss}
\end{align}
where $s$ is the usual Mandelstem variable and the decay rate for $H_2\rightarrow \chi\chi$ and $H_2\rightarrow SS$ are expressed by:
\begin{align}
\Gamma(H_2\rightarrow
\chi\chi)&=\frac{g^2M_{H_2}}{2\pi}(1-\frac{4M^2_{\chi}}{M^2_{H_2}})^{3/2},
\label{decay1}\\
\Gamma(H_2\rightarrow SS)&=\frac{\lambda^2_{\phi s}M^2_{\chi}}{16\pi
	g^2M_{H_2}}(1-\frac{4M^2_S}{M^2_{H_2}})^{1/2}. \label{decay2}
\end{align}

In the following, we calculate self-interaction for the case of Dirac fermionic DM which includes processes $\chi\chi\longrightarrow \chi\chi$ and $\chi\chi\longrightarrow SS$. The cross section of these processes are given by:
\begin{align}
\sigma(\chi\chi\rightarrow \chi\chi)/M_{\chi}&=\frac{g^2s}{16\pi M_{\chi}}(1-\frac{4M_{\chi}^2}{s})^{2}|\frac{1}{\sqrt{1+(\nu_1g/M_{\chi})^2}}\times\frac{1}{s-M_{H_2}^2+iM_{H_2}\Gamma_{H_2}}\nonumber\\&-\frac{g\nu_1/M_{\chi}}{\sqrt{1+(\nu_1g/M_{\chi})^2}}\times\frac{1}{s-M_{H_1}^2+iM_{H_1}\Gamma_{H_1}}|^2,\label{xxxx}
\end{align}
\begin{align}
\sigma(\chi\chi\rightarrow SS)/M_{\chi}&=\frac{\lambda^2_{\phi s}g^2M_{\chi}}{32\pi }(1-\frac{4M_{s}^2}{s})^{1/2}(1-\frac{4M_{\chi}^2}{s})^{1/2}
\nonumber\\&\times|\frac{2M_{\chi}/g}{1+(\nu_1g/M_{\chi})^2}\frac{1}{s-M_{H_2}^2+iM_{H_2}\Gamma_{H_2}}\nonumber\\&+\frac{2\nu^2_1g/M_{\chi}}{1+(\nu_1g/M_{\chi})^2}\frac{1}{s-M_{H_1}^2+iM_{H_1}\Gamma_{H_1}}|^2
\label{xxss}.
\end{align}

We also calculate the DM self scattering cross-sections for processes   $S\chi\longrightarrow S\chi$ in non-relativistic limit. The cross sections can be written as:  
\begin{align}
\sigma(S\chi\rightarrow S\chi)/\overline{M}&\simeq\frac{\lambda^2_{\phi s}M_{\chi}^4
	}{8\pi p^2(M_{\chi}+M_s)^3}[(\frac{1}{(m_{H_2}^2)}-\frac{1}{(4p^2+m_{H_2}^2)})(\frac{2M_{\chi}/g}{1+(\nu_1g/M_{\chi})^2})^2\nonumber\\&+(\frac{1}{(m_{H_1}^2)}-\frac{1}{(4p^2+m_{H_1}^2)})(\frac{2\nu^2_1g/M_{\chi}}{1+(\nu_1g/M_{\chi})^2})^2]\nonumber\\
	&\times[(\frac{E_p^2+m^2_{\chi}}{\sqrt{E_p^2-m^2_{\chi}}}\sqrt{E_k^2-m^2_{\chi}}].\label{xs}
\end{align}
where $p$, $E_p$ and $E_k$ are  momentum of initial fermionic DM, energy of initial fermionic DM and energy of final fermionic DM, respectively.

We also calculate the following formula for decay rates of $H_1\rightarrow \chi\chi$, $H_1\rightarrow SS$ and $H_1\rightarrow H_2H_2$:
\begin{align}
&\Gamma(H_{1}\rightarrow \chi\chi)=\frac{M  _{H_{1}}a_{H_{1}\chi\chi}^2}{2\pi}(1-\frac{4M^2_{\chi}}{M^2_{H_{1}}})^{3/2}, \label{decay1}\\
&\Gamma(H_{1}\rightarrow SS)=\frac{a_{H1SS}^2}{16\pi
M_{H_{1}}}(1-\frac{4M^2_s}{M^2_{H_{1}}})^{1/2},\\
&\Gamma(H_{1}\rightarrow H_2H_2)=\frac{a_{H_1H_2H_2}^2}{16\pi
M_{H_{1}}}(1-\frac{4M^2_{H_2}}{M^2_{H_{1}}})^{1/2},\label{decay2}&
\end{align}
where 
\begin{eqnarray}
& a_{H_1\chi\chi}=\frac{g^2\nu_1
}{\sqrt{(g^2\nu_1^2+M_{\chi}^2)}},\nonumber\\
& a_{H1SS}=\frac{2\nu_1\lambda_{\phi
s}}{(1+(\nu_1g/M_{\chi})^2)^{1/2}}+\frac{(M^2_s-2\lambda_{\phi s}M^2_{\chi}/g^2)}{\nu_1\sqrt{1+(\nu_1g/M_{\chi})^2}},\nonumber\\
& a_{H1H_2H_2}=\frac{M^2_{H_1}}{2(1+(\nu_1g/M_{\chi})^2)^{5/2}}[\nu_1^4(\frac{g}{M_{\chi}})^5-\frac{g}{M_{\chi}}]\nonumber.
\end{eqnarray}

\end{document}